\documentclass[a4paper,11pt]{article}
\usepackage{jheppub} 
\usepackage{lineno}

\title{\boldmath Research on the application of loop quantum theory model in black hole quantum information}

\author{Yangting Liu}
\affiliation{Xi'an Jiaotong University,\\
No.28 Xianning West Road, Xi'an, Shaanxi, P.R.China}

\emailAdd{18954559536@stu.xjtu.edu.cn}

\abstract{An important reason why it is currently difficult to unify relativity theory and quantum theory is the quantum information paradox. The information engulfment pointed out by general relativity violates the principles of quantum mechanics. An important reason why the industry does not have a clear understanding of this phenomenon is the current lack of a theoretically solvable cosmological model. Based on the complete model of loop quantum theory, this article solves different levels of Hamiltonian constraint models and simulates black hole information transfer dynamics, especially at extreme points, from analytical results to step-by-step quantum corrections, and attempts to compare the performance of different physical models in simulating quantum Advantages during information transmission. Our study shows that even second-order expansions are sufficient to distinguish differences in dynamics at the black hole extremes, but to truly identify a model that has the potential to describe quantum information transfer mechanisms and is significantly different from other models, the theoretical analytical solution should at least extend to Level three and above. In addition, the research results such as computational simulation methods and related conclusions cited and improved in this article can provide certain theoretical support and new insights for the research prospects of general relativity loop quantum cosmology and the intersection of quantum information and quantum fields.}


\begin{document}
\maketitle
\flushbottom

\section{Introduction}
\label{sec:intro}

{C}{lassical} general relativity (GR) theory has huge limitations at singularities. At these locations, physical quantities such as curvature and density of space become infinite, which will cause the relevant equations of general relativity to fail and make it impossible to predict the dynamics of the system. feature. Several newly developed theories address these issues, such as string theory and quantum theory. A complete quantum theory used to solve the singularity of general relativity is called loop quantum gravity theory, which is a concept based on loop quantum cosmology ~\cite{ref1}. Under this theoretical system, the origin of the black hole (the starting point of the Big Bang) is considered to be a big bounce that replaces the singularity, connecting the classical universe and the expanding universe with a space with great curvature but limited space. \cite{ref2,ref3} In the Big Bounce model, instead of starting from an infinitely dense point, the universe starts from a highly compressed but finitely dense state and then begins to expand. Based on this basic theory of loop quantum gravity (LQC) \cite{ref4}, many new scholars have devoted themselves to developing a new and more accurate theory in recent years. From the most basic, the Friedmann-Lemaître-Robertson-Volkor (FLRW) model \cite{ref5}, which describes the isotropy and uniformity of the universe, is governed by the Friedmann equations, which are specific versions of Einstein's field equations. The solution takes into account the effects of matter, radiation and the cosmological constant (dark energy). To the latest proposed black hole model with axial symmetry based on the Kantosky-Sachs cosmological model. From the most basic, the Friedmann-Lemaître-Robertson-Volkor (FLRW) model, which describes the isotropy and uniformity of the universe, is governed by the Friedmann equations, which are specific versions of Einstein's field equations. The solution takes into account the effects of matter, radiation and the cosmological constant (dark energy). The latest proposed black hole model with axial symmetry is based on the cosmological model of Kantosky Sachs. These models all qualitatively solve the problem of singularities. Their explanation is that the singularity is substituted by a hypersurface akin to a nearby space jump \cite{ref6}.In the temporal framework, the region of gravitational capture, identified as a black hole, precedes an antithetical region characterized as a white hole. This sequence facilitates the replacement of the singularity with a progressive evolution from a black hole phase to a white hole phase. More accurate models of loop quantum theory \cite{ref7,ref8} are based on aggregation of non-uniform solutions of a more general class and then reduction to the uniform case while preserving covariance, i.e. preserving the absence of anomalies in the effective constraint algebra. At the same time, it is gratifying that this model also reveals some new internal phenomena of black holes, such as the internal horizon and the existence of Euclidean space-time regions inside black holes.

Subsequent physical models used to describe black holes are all similar \cite{ref9,ref10,ref11,ref12,ref13,ref14}. However, some processing details are different when using semi-classical methods to reduce the degree of freedom of the system (such as studying spherically symmetric black hole models). An alternative approach involves formulating a quantum Hamiltonian within the Loop Quantum Cosmology (LQC) framework, deriving its foundation from the principles of loop quantum gravity. Subsequent physical models used to describe black holes are all similar. However, some processing details are different when using semi-classical methods to reduce the degree of freedom of the system (such as studying spherically symmetric black hole models). Another processing method is to construct a quantum Hamiltonian in the LQC model based on ring quantum gravity. Although this scheme of gravity quantization has been speculated to have excellent properties in smoothing singularities and solving quantum information paradoxes, this idea has not been taken seriously by the industry. The main reason is that in the analysis of some classical black hole dynamics, the results of quantum processing are no different from those of classical models \cite{ref15}.

The document \cite{ref15} compiles the Hamiltonian describing the black hole model under different models. Includes the classical Hamiltonian based on the Kantowski-Sachs metric of the Ashteka-Barbero variable, the Hamiltonian using the LQC classical regularization method and the Hamiltonian obtained from the ring quantum theory LQG, and As a result, the differences in the internal dynamic evolution of black holes under different physical models were obtained through numerical calculations. And when the initial conditions are taken at the black hole event horizon, the conditions before the big rebound of the universe get significant differences in the numerical simulation results. This article uses a similar research idea, but the focus is no longer on the simple description of the internal dynamics of the black hole, but also on the transmission of quantum information inside the black hole under different physical models. It is worth noting that the above models are all based on the derivation results given by the basic general theory of relativity or the ring quantum gravity theory. There are also recent studies \cite{ref16} that have further improved the spherically symmetric black hole model. By reducing the expression method of the phase space and reducing the spherical symmetry, a 1+1-dimensional field theory model with infinite degrees of freedom is obtained. The advantage of these improved models is that the parameterized gravity term is eliminated through purely mathematical methods, which greatly facilitates quantum information encoding and computer simulation.

The information encoding and transmission mechanism in black holes, especially under extreme conditions, is currently unclear. Hawking pointed out that the traditional concept is incorrect, and black holes can transmit information through the "Hawking radiation" effect. Black holes follow the laws of thermodynamics, and black hole entropy is proportional to the event horizon area \cite{ref17,ref18}. According to Hawking radiation theory, in the quantum fluctuation region near the black hole event horizon, particle pairs (a particle and an antiparticle) can be spontaneously produced. Typically, these pairs of particles immediately annihilate each other. But near the event horizon, one particle may be sucked into the black hole while the other escapes into outer space. According to thermodynamic theory, these escaping particles carry no information, as if the incoming information was annihilated in the black hole. Viewed in this way, this contradicts the unitarity required by quantum mechanics \cite{ref19,ref20}. This information paradox is an important obstacle that makes it difficult to unify the current quantum mechanics theory and the theory of relativity. Although new research results provide some solutions to the information paradox, such as the spontaneous random collapse of modified quantum states\cite{ref19}, the explanation from the encoding and decoding mechanism of information inside the black hole through the principle of complementarity\cite{ ref20,ref21}, this article is based on the above research theory, re-derivating a more complete black hole quantum gravity model, and tracking the behavior of quantum information inside the black hole, especially in the extreme region, through computer simulation. This provides a new theoretical framework explanation for the capture and release mechanism of information in black holes, and may provide support for further explanations of information paradoxes and understanding of the nature of quantum information entropy in the future.

The research content of this paper is organized into the following sections: Section I briefly reviews and derives a comprehensive black hole model within the framework of loop quantum theory and demonstrates its suitability for computer simulations. Section II introduces a novel method for information transmission and decoding based on the newly established theoretical model. In Section III, we employ computers to simulate the dynamics of quantum information in black holes, particularly focusing on the behavior near extremal points, using the new model and method. Section IV presents our conclusions, where we propose a potential solution to the quantum information paradox based on the results of our computer simulations. This proposal also highlights the potential contributions of our findings to enhancing the quantum theory of gravity and deepening our fundamental understanding of black holes.

\section{Loop Quantum Theory Black Hole Model}

\subsection{LQG semi-classical cosmological model and its quantization}

The semiclassical model of loop quantum theory is derived from the approach taken to uniform isotropic cosmology \cite{ref25,ref26}. Initially, one must contemplate a semiclassical coherent state within the Hilbert space as per the ring quantum gravity theory, followed by computing the expected value of the ring quantum Hamiltonian operator in relation to this coherent state. The result of its semi-classical expansion is the Hamiltonian in the effective theory. The core essence is to construct a Hamiltonian used to describe the black hole system. LQG theory points out that this Hamiltonian may have the following form:$\displaystyle \widehat{H[N]}=\widehat{H_{E}[N]}+(1+\beta^{2})\widehat{H_{L}[N]}$, The first term is called "Euclidean" and the second term is called "Lorentzian". Due to the complexity of its analysis and the uncertainty of the spectrum, it is currently difficult to provide analytical results theoretically, and numerical calculations face the problem of excessive computational complexity \cite{ref27,ref28}. But we can use $\mu_0$ theory to linearly expand the Hamiltonian in isotropic and continuous coherent states - as can be seen from the following theoretical analysis, this is true whether it is for the volume operator $\hat V$ that is difficult to analyze Either the Lorentz Hamiltonian or the Lorentz Hamiltonian is applicable, and comparing the terms of the expansion with the results in the literature\cite{ref15} it is not difficult to see that expansions at all levels (especially the first and second level expansions) are precisely through the classical generalized The Schwinger model of relativity uses different methods for regularization and quantization.

As mentioned before, the current theory is limited to the analytical calculation of the Hamiltonian operator in the classical limit, so the Hamiltonian operator given by ring quantum theory:

\begin{equation}
\label{deqn_ex1a}
\begin{gathered}
\hat{H}[N,\epsilon]=\sum_{\alpha}N(\nu_{\alpha})\epsilon^{mnp}\mathrm{Tr}\Big(\Big(h_{\partial P_{m}(\epsilon)}-h_{\partial P_{m}(\epsilon)}^{-1}\Big)h_{p}^{-1}[h_{p},\hat{V}]\Big) \\
-\frac{1}{2}(1+\gamma^{2})\sum_{\alpha}N(\nu_{\alpha})\epsilon^{mnp}\mathrm{Tr}\Big(h_{m}^{-1}[h_{m},\bar{K}]h_{n}^{-1}[h_{n},\bar{K}]h_{p}^{-1}[h_{p},\hat{V}]\Big) 
\end{gathered}
\end{equation}

Although expressed in the form of a lag function $N$ and a coordinate space parameter $\epsilon$, this small quantity does not appear in the explicit expression, and its limit is found in the dual space of the Hamiltonian: $\displaystyle \langle\hat{H}^{*}[N]{\mathcal X}\mid\Psi\rangle=\lim_{\epsilon\rightarrow0}\langle{\mathcal X}\mid\hat{\mathcal H}[N,\epsilon]\Psi\rangle $ Unfortunately, this is an expression obtained in the case of the weak limit, which is not consistent with the strong limit theory in quantum field theory. For details on how to use a holomorphic expression to relate the Ashtekar coordinates $(A,E)$, the field strength $F_{mn}$ and the external curvature $K$, the graph method of the quasi-lattice gauge theory used can be found in \cite{ref15}.

The first difficulty that needs to be dealt with in this formula is the processing of operator $\hat{V}_{v}=\sqrt{|\hat{Q}_{v}|}$. Since it is an expression of a square root, the entire Hamiltonian operator no longer has the properties of a polynomial. Therefore, refer to the expression of \cite{ref29} and perform a semi-classical expansion:

\begin{equation}
\label{deqn_ex1a}
\hat V_{GT}^{(v)} = {\langle {\hat Q_v}\rangle ^{2q}}[1 + \sum\limits_{n = 1}^{2k + 1} {{{( - 1)}^{n + 1}}} \frac{{q(1 - q) \cdots (n - 1 - q)}}{{n!}}{\text{ }}\left. { \times {{\left( {\frac{{\hat Q_v^2}}{{{{\langle {{\hat Q}_v}\rangle }^2}}} - 1} \right)}^n}} \right] + O({\hbar ^{k + 1}})
\end{equation}

The results of $\hat{Q_v}$ quantization are shown below:

\begin{equation}
\label{deqn_ex1a}
{\hat Q_v} =  - i{(\beta {a^2})^3}{\varepsilon _{\alpha \beta \gamma }}\frac{{\hat p_s^\alpha (e_x^ + ) - \hat p_t^\alpha (e_x^ - )}}{2} \times \frac{{\hat p_s^\beta (e_y^ + ) - \hat p_s^\beta (e_y^ - )}}{2}\frac{{\hat p_s^\gamma (e_z^ + ) - \hat p_s^\gamma (e_z^ - )}}{2}
\end{equation}

Within this context, \( p_s^i \) represents the gauge-covariant fluxes across the two-dimensional faces \( S_e \) situated within the dual lattices denoted as \( \gamma^* \).
 For specific processing details, see \cite{ref30}. Only the final expression is shown below:

\begin{equation}
\label{deqn_ex1a}
\hat{p}_{v}^{\pm1}(e):=\mp\frac{1}{\sqrt{2}}(\hat{p}_{v}^{x}(e)\pm i\hat{p}_{v}^{\mathrm{v}}(e)),\quad\hat{p}_{v}^{0}(e)=\hat{p}_{v}^{z}(e)
\end{equation}

With $v=s,t$. After the expression of the volume operator is determined, explicit expressions for the two parts of the Hamiltonian operator can be determined: the "Euclidean" part and the "Lorentzian" part:

\begin{equation}
\label{deqn_ex1a}
\widehat{H_{E}[N]}=\frac{1}{i\beta a^{2}t}\sum_{v\in V(\gamma)}N(\nu)\cdot\sum_{e_{I},e_{J},e_{K}\mathrm{~at~}\nu}\epsilon^{IJK}\mathrm{tr}(h_{a_{l}}h_{e_{K}}[\hat{V}_{\nu},h_{e_{K}}^{-1}])
\end{equation}

Consider the edges \( e_I, e_J, \) and \( e_K \), each oriented to emanate from the vertex \( v \). We define \( \epsilon_{IJK} \) as the sign of the determinant formed by the wedge product \( e_I \wedge e_J \wedge e_K \), denoted as \( \text{sgn}[\text{det}(e_I \wedge e_J \wedge e_K)] \). Here, \( \alpha_{IJ} \) represents the minimal loop encircling a plaquette constituted by \( e_I \) and \( e_J \), which extends outward along \( e_I \) and returns via \( e_J \), with \( v \) serving as its terminal point. Employing a similar conceptual framework, the Lorentzian component is articulated as follows:

\begin{equation}
\label{deqn_ex1a}
\begin{aligned}
\widehat{H_{L}[N]}& =\frac{-1}{2i\beta^{7}a^{10}t^{5}}\sum_{v}N(v)  \times\sum_{e_{I},e_{J},e_{K}\mathrm{~at~}v}\varepsilon^{IJK}\mathrm{tr}([h_{e_{I}},[\hat{V},\hat{H}_{E}]]h_{e_{I}}^{-1} \\
&\times[h_{e_{J}},[\hat{V},\hat{H}_{E}]]h_{e_{J}}^{-1}[h_{e_{K}},\hat{V}_{v}]h_{e_{K}}^{-1}).
\end{aligned}
\end{equation}

Another key issue is the choice of hysteresis function N. A common selection method is $N=\mathrm{sgn}(p_{b})\sqrt{|p_{a}|})$, because this approach can make the expected value of the classical Hamiltonian of general relativity analytically solvable \cite{ref15}. Expanding this result directly using a computer will generate a calculation amount of $10^19$ degrees of freedom, which is beyond our ability to bear, so we use existing numerical discretization methods to simplify it. The first step is to write an arbitrary operator in the form of a flux:

\begin{equation}
\label{deqn_ex1a}
\begin{gathered}
\begin{gathered}
  \hat O = {{\hat O}_1}{{\hat O}_2} \cdots {{\hat O}_m} =  \hfill \\
  \left( \begin{gathered}
  {\text{with: }}[{D^\prime }({h_e}),{D^\prime }({h_{{e^\prime }}})] = 0 = [\hat p_s^i(e),p_t^j({e^\prime })], \hfill \\
  [\hat p_s^i(e),\hat p_s^j({e^\prime })] =  - it{\delta _{e{e^\prime }}}{\epsilon_{ijk}}\hat p_s^k(e), \hfill \\
  [\hat p_t^i(e),\hat p_t^j(e')] =  - it{\delta _{e{e^\prime }}}{\epsilon_{ijk}}\hat p_t^k(e), \hfill \\
  [\hat p_s^i(e),{D^i}({h_{{e^\prime }}})] = it{\delta _{e{e^\prime }}}{D^{\prime \imath }}({\tau ^i}){D^\imath }({h_e}), \hfill \\
  [\hat p_t^i(e),{D^i}({h_{{e^\prime }}})] =  - it{\delta _{e{e^\prime }}}{D^\imath }({h_e}){D^{\prime \imath }}({\tau ^i}), \hfill \\ 
\end{gathered}  \right) \hfill \\
  {\text{  }} = {{\hat O}_2} \cdots {{\hat O}_m}{{\hat O}_1} + \sum\limits_{k = 1}^{m - 1} {\sum\limits_{{\mathcal{I}_k}} {\left( {\prod\limits_{l \in \mathcal{I} - {\mathcal{I}_k}} {{{\hat O}_l}} } \right)} } [[ \cdots [[{{\hat O}_1},{{\hat O}_{{i_1}}}],{{\hat O}_{{i_2}}}] \cdots ],{{\hat O}_{{i_k}}}] \hfill \\
  {\text{    = }} \cdots [[D_{ab}^l({h_e}),\hat p_t^{{\alpha _1}}(e)],\hat p_t^{{\alpha _2}}(e)] \cdots \hat p_t^{{\alpha _m}}(e)] \hfill \\
  {\text{  }} = {(it)^m}D_{a{b_m}}^l({h_e})D_{{b_m}{b_{m - 1}}}^{\prime \prime }({\tau ^{{\alpha _m}}})D_{{b_{m - 1}}{b_{m - 2}}}^{\prime \prime }({\tau ^{{\alpha _{m - 1}}}}) \cdots D_{{b_1}b}^{\prime \prime }({\tau ^{{\alpha _1}}}) \hfill \\
  {\text{    = }}\left( {with:{\text{ equal(1) in appendix A}}} \right) \left( {\prod\limits_{i = 1}^m {\hat p_s^{{\alpha _i}}} (e)} \right)\left( {\prod\limits_{j = 1}^n {\hat p_t^{{\beta _j}}} (e)} \right)D_{ab}^i({h_e}) \hfill \\ 
\end{gathered} 
\end{gathered} 
\end{equation}

Then generalize the field operator. The result is shown in equ(8). The expressions for some of these integrals are as follows:

$$
\begin{aligned}
I_1& =\int_{-\infty}^{\infty}\mathrm{d}xe^{-ax^{2}+bx}\frac{\sinh(x\eta)}{x}=\frac{\pi}{2}\left(\mathrm{erfi}\left(\frac{b+\eta}{2\sqrt{a}}\right)-\mathrm{erfi}\left(\frac{b-\eta}{2\sqrt{a}}\right)\right)
\end{aligned}
$$

and

$$
I_{2}=\int_{-\infty}^{\infty}\mathrm{d}xe^{-ax^{2}+bx}\mathrm{pol}(x,\partial_{z})\left.\frac{e^{\pm\eta x}}{f(z)}\right|_{z=\eta}
$$

\begin{figure*}
\begin{equation}
\label{deqn_ex1a}
\begin{gathered}
  \left( \begin{gathered}
  {\text{with }}{\langle \hat p_s^{{\alpha _1}}(e) \cdots \hat p_s^{{\alpha _m}}(e)\hat p_t^{{\beta _1}}(e) \cdots \hat p_t^{{\beta _n}}(e)D_{ab}^\prime ({h_e})\rangle _{{z_e}}} \hfill \\
  {\text{        }} = {( - 1)^n}{e^{ - ({\beta _1} +  \cdots  + {\beta _n})\overline {{z_e}} }}\langle \hat p_s^{{\beta _n}}(e) \cdots \hat p_s^{{\beta _1}}(e)\hat p_s^{{\alpha _1}}(e) \cdots  \hfill \\ 
\end{gathered}  \right) \hfill \\
   = \hat F_{\imath ab}^{{\alpha _1} \cdot  \cdot  \cdot {\alpha _m}} = \hat p_s^{{\alpha _1}}(e) \cdots \hat p_s^{{\alpha _m}}(e)D_{ab}^l({h_e}) 
   = {\langle {{\hat F}^{{n_1} \cdots {\alpha _n}}}\rangle _{{z_e}}}  \hfill \\ =\delta (\sum\limits_i {{\alpha _i}} ,0){t^m}\prod\limits_{i = 1}^m {\frac{1}{{{{(1 + |{\alpha _i}|)}^{1/2}}}}} {e^{t/4}}\int_{ - \infty }^\infty  {\text{d}} xx\prod\limits_{k = 1}^m ( \frac{{{a_k} - 1}}{2}x - \frac{{{\partial _y}}}{2} + \sum\limits_{i = 1}^k {{\alpha _i}}  - \frac{{{a_k}}}{2}) \hfill \\ \frac{{{e^{ - \frac{t}{4}{x^2} + x\eta }}}}{{2\sinh (y)}}{|_{y \to \eta }} + O({t^\infty }) \hfill \\
   = {t^m}{e^{b{z_e}}}\sum\limits_{\frac{{0 \leqslant d \leqslant t}}{{d + t \in \mathbb{Z}}}} {\frac{{2 - \delta (d,0)}}{2}} {e^{ - \frac{t}{4}(2{d^2} - 1)}}\int {\text{d}} x{e^{ - \frac{1}{4}t({x^2} - 2dx)}}{F_l}\left( {\frac{{x - 1}}{2} - d,\frac{{x - 1}}{2},\frac{{{\partial _\eta }}}{2}} \right) \cdot \hfill \\ \frac{{\sinh (x\eta )}}{{\sinh (\eta )}} + O({t^{ - \infty }}), \hfill \\
   = {t^m}{e^{b{z_e}}}\sum\limits_{\frac{{0 \leqslant d \leqslant t}}{{d + t \in \mathbb{Z}}}} {\frac{{2 - \delta (d,0)}}{2}} {e^{ - \frac{t}{4}(2{d^2} - 1)}}\int {\text{d}} x{e^{ - \frac{1}{4}t({x^2} - 2dx)}}\delta (\sum\limits_{i = 1}^m {{\alpha _i}}  - a + b,0) \cdot  \hfill \\
  {\text{   }}\prod\limits_{i = 1}^m {\frac{1}{{{{(1 + |{\alpha _i}|)}^{1/2}}}}} {(\frac{1}{{(\imath  + a)!(\imath  - a)!(\imath  + b)!(\imath  - b)!}})^{1/2}} \times \frac{{{{( - 1)}^{d - 2a + b}}x(x - 2d)}}{{{{(x - d + \imath )}_{2i + 1}}}} \hfill \\ \prod\limits_{k = 1}^m {\left( {{\alpha _i}\frac{{x - 1}}{2} - \frac{{{\partial _\eta }}}{2} + \sum\limits_{i = 1}^k {{\alpha _i}} } \right)}    \hfill \\
  \times \sum\limits_{z = 0}^{t + d} {\left[ {\frac{{{{( - 1)}^z}{{(\imath  + d)}_z}{{(\frac{{x - 1}}{2} + \frac{{{\partial _q}}}{2} + b - z + \imath )}_{\imath  + a}}}}{{z!}} \cdot } \right.} \left. {\frac{{{{\left( {\frac{{x - 1}}{2} - \frac{{{\partial _q}}}{2} - d - b + z} \right)}_{\imath  - a}}}}{{z!}}} \right] \hfill \\
   \times \sum\limits_{z = 0}^{i - d} {\left[ {\frac{{{{( - 1)}^z}{{(\imath  - d)}_z}{{(\frac{{x - 1}}{2} + \frac{{{\partial _n}}}{2} - d - z + \imath )}_{\imath  - b}}}}{{z!}}} \right.} \left. {\frac{{{{(\frac{{x - 1}}{2} - \frac{{{\partial _n}}}{2} + z)}_{\imath  + b}}}}{{z!}}} \right]\frac{{\sinh (x\eta )}}{{\sinh (\eta )}} + O({t^{ - \infty }}) \hfill \\ 
\end{gathered}   
\end{equation}
\end{figure*}

Let \( a > 0 \), \( b \in \mathbb{R} \), and \( f \) be a given function. Consider the polynomial \( \text{pol}(x; \partial z) \), which is a polynomial in \( x \) and \( \partial z \). This is the first step of computing, currently not considering the implementation of their concrete forms. Since \( I_1 \) as a function of \( a \) and \( b \) is known, the next step of the computing is to compute \( I_2 \). To do this, we first expand \( \text{pol}(x; \partial z) \) and write the integrand of \( I_2 \) as a linear combination of \( (\partial^n z f(z)) x^m e^{-ax^2 + bx} \eta x \). Then, by substituting the results of \( \int dx\, x^n e^{-ax^2 + bx} \eta x \), \( I_2 \) can be easily computed.

Eliminate the flux terms in the operator and keep only the polynomial terms, which is simplified as follows, as the equation(9) shows. And finally we can use this result to get the spectrum of the field quantity:

$$
\begin{gathered}
  D_{{a_1}{b_1}}^\imath ({h_e}){[\hat p_s^0(e)]^{{m_1}}}{[\hat p_t^0(e)]^{{n_1}}} 
   \cdots D_{{a_k}{b_k}}^\imath ({h_e}){[\hat p_s^0(e)]^{{m_k}}}{[\hat p_t^0(e)]^{{n_k}}} \hfill \\ 
\end{gathered} 
$$

And finally this result can be used to obtain the spectrum of the field quantity. The result of applying the operator to the ground state is shown in equation (10). Before this, the coherent state needs to be defined: $|\Psi_{\vec{g}}\rangle=\bigotimes_{e\in E(\gamma)}|\psi_{g_{e}}\rangle $. The matrix element of the operator acting on the coherent state is determined by the following formula: 

$$\langle\psi_{g_{\epsilon}}|\hat{O}_{i}|\psi_{g_{\epsilon}^{\prime}}\rangle=\langle\psi_{g_{\epsilon}}|\psi_{g_{\epsilon}^{\prime}}\rangle(E_{0}(g_{e},g_{e}^{\prime})+tE_{1}(g_{e},g_{e}^{\prime})+O(t^{\infty}))$$

\begin{figure*}
\begin{equation}
\label{deqn_ex1a}
\begin{gathered}
  D_{{a_1}{b_1}}^\imath ({h_e}){[\hat p_s^0(e)]^{{m_1}}}{[\hat p_t^0(e)]^{{n_1}}} \cdots D_{{a_k}{b_k}}^\imath ({h_e}){[\hat p_s^0(e)]^{{m_k}}}{[\hat p_t^0(e)]^{{n_k}}} 
   = {[\hat p_s^0(e)]^{\sum\limits_{i = 1}^k {{m_i}} }}{[\hat p_i^0(e)]^{\sum\limits_{i = 1}^k {{n_i}} }}\hfill \\ \prod\limits_{i = 1}^k {D_{{a_i}{b_i}}^\imath } ({h_e}) - t\left( {\sum\limits_{i = 1}^k {{a_i}} \left[ {\sum\limits_{l = i}^k {{m_i}} } \right]{{[\hat p_s^0(e)]}^{(\sum\limits_{i = 1}^k {{m_i}} ) - 1}}{{[\hat p_t^0(e)]}^{\sum\limits_{i = 1}^k {{n_l}} }}\prod\limits_{i = 1}^k {D_{{a_i}{b_i}}^\imath } ({h_e})} \right) \hfill \\
   + t(\sum\limits_{i = 1}^k {{b_i}} [\sum\limits_{l = i}^k {{n_i}} ]{[\hat p_s^0(e)]^{\sum\limits_{i = 1}^k {{m_i}} }}{[\hat p_t^0(e)]^{(\sum\limits_{i = 1}^k {{n_i}} ) - 1}}\prod\limits_{i = 1}^k {D_{a,{b_i}}^\iota } ({h_e})) + O({t^2}) \hfill \\
   = {( - 1)^{\sum\limits_{i = 1}^k {{n_i}} }}{[\hat p_s^0(e)]^{\sum\limits_{i = 1}^k {({m_i} + {n_i})} }}\prod\limits_{i = 1}^k {D_{{a_i}{b_i}}^i} ({h_e}) - t{( - 1)^{\sum\limits_{i = 1}^k {{n_i}} }}\left( {\sum\limits_{i = 1}^k {{a_i}} \left[ {\sum\limits_{l = i}^k {{m_l}} } \right] + \sum\limits_{i = 1}^k {{b_i}} \left[ {\sum\limits_{l = i}^k {{n_l}} } \right]} \right) \hfill \\
   \times {[\hat p_s^0(e)]^{\sum\limits_{i = 1}^k {({m_i} + {n_i})}  - 1}}\prod\limits_{i = 1}^k {D_{{a_i}{b_i}}^t} ({h_e}) \hfill \\ 
\end{gathered} 
\end{equation}
\end{figure*}

With this result, the two parts of the Hamiltonian constraint can be expanded and simplified. The specific results are shown in the next section.

\subsection{Expected value of Hamiltonian}

Equations (2.8) and (2.9) provide the method for calculating the Lorentz Hamiltonian and analytically expanded Euclidean Hamiltonian in this part. Recalling the derivation of $\langle\hat{F}_{\textit{la}b}^{\alpha_1\cdots\alpha_m}\rangle_{z_e}$, we get:

\begin{equation}
\label{deqn_ex1a}
\begin{gathered}
  {\langle D_{ab}^l(N({h_e}))\rangle _{{z_e}}} = {\left\langle {N({h_e})} \right\rangle _{{z_e}}}({g_0} + tN({h_e}){g_1}(\eta ) + O({t^2})) \hfill \\
  {\text{                       }} = f(t)\frac{{{e^{\frac{{{\pi ^2}}}{t}}}\eta }}{{\sinh (\eta )}}\left( {{g_0} + tN({h_e}){g_1}(\eta ) + O({t^2})} \right) \hfill \\ 
\end{gathered} 
\end{equation}

Among them: $N(h_e)$ represents the hysteresis function in the Hamiltonian model of general relativity. Therefore, the expected value of the ground state can be calculated as follows:

\begin{equation}
\label{deqn_ex1a}
\begin{gathered}
  {\left\langle {{{(\hat p_s^0(e))}^m}D_{ab}^\iota ({h_e})} \right\rangle _{{z_e}}} = {e^{b\eta }}{\left( { - t\frac{{{\partial _\eta }}}{2}} \right)^m}{e^{ - b\eta }} < D_{ab}^\iota ({h_e}){ > _{{z_e}}} \hfill \\
   = {\langle N({h_e})\rangle _{{z_e}}}{( - \eta )^m}[{g_0} + tN({h_e}){g_l}(\eta )] + {\langle N({h_e})\rangle _{{z_e}}}\frac{{m(m + 1)}}{4}{( - \eta )^{m - 2}}{g_0}t \hfill \\
  {\text{  }} + {\langle N({h_e})\rangle _{{z_e}}}\frac{m}{2}{( - \eta )^{m - 1}}(\coth (\eta ) + bN({h_e}))N({h_e}){g_0}t + O({t^2}) \hfill \\ 
\end{gathered} 
\end{equation}

The results of this part are obtained based on the calculation results in (A). Based on this, if the operator is further written in the following form: $\displaystyle \sum_{\vec{\alpha}}\mathcal{T}^{\alpha_1\alpha_2\cdots\alpha_m}\hat{O}_{\alpha_1\alpha_2\cdots\alpha_m}$, The complete expression of its flux is:

\begin{equation}
\label{deqn_ex1a}
\begin{aligned}
\left\langle\mathcal{M}\right\rangle_{z_{e}}& \cong(\langle\hat{p}_{s}^{0}(e)\rangle_{z_{e}})^{N_{0,s}}(\langle\hat{p}_{t}^{0}(e)\rangle_{z_{e}})^{N_{0,t}}  \\
&\times(\langle D_{\frac{11}{22}}^{\frac{1}{2}}(h_{e})\rangle_{z_{e}})^{M_{0,+}}(\langle D_{-\frac{1}{2}-\frac{1}{2}}^{\frac{1}{2}}(h_{e})\rangle_{z_{e}})^{M_{0,-}}\langle\mathcal{M}^{\prime}\rangle_{z_{e}}
\end{aligned}
\end{equation}

Consider \( M \) as an arbitrary monomial composed of holonomies and fluxes. Define \( M_0 \) as the resultant operator obtained by excluding all instances of \( \hat{p}_0(e) \) and \( D_{aa}^{1/2}(h_e) \) from \( M \). The quantities \( N_{0,s} \) and \( N_{0,t} \) represent the counts of \( \hat{p}_{s}^0(e) \) and \( \hat{p}_{t}^0(e) \) within \( M \), respectively, while \( M_{0+} \) and \( M_{0-} \) denote the counts of \( D_{1/2,1/2}^{1/2}(h_e) \) and \( D_{-1/2,-1/2}^{1/2}(h_e) \), respectively. The leading order of \( \langle M \rangle_{\text{ze}} \) is precisely \( O(t^{M_+ + N_+}) \) if and only if the leading order of \( \langle M_0 \rangle_{\text{ize}} \) is exactly \( O(t^{M_+ + N_+}) \). Here, \( N_{\pm} \) signifies the count of \( \hat{p}^{\pm 1}(e) \), and \( M_+ \) (respectively \( M_- \)) corresponds to the number of \( D_{1/2,1/2}^{1/2}(h_e) \) [respectively \( D_{-1/2,-1/2}^{1/2}(h_e) \)].

Rearrange the expected value simplification of the two-part Hamiltonian as equation(2.13) shows and the simplified result of the Lorentz term is as follows in equation (2.14) shows.

\begin{figure*}
\begin{equation}
\label{deqn_ex1a}
\begin{gathered}
  \hat H_E^{(n)}(v) = \sum\limits_{{e_I},{e_J},{e_K}} {\hat H_E^{(n)}} (v;{e_I},{e_J},{e_K}) = \sum\limits_{{e_I},{e_J},{e_K}} {\frac{1}{{i\beta {a^2}t}}{\varepsilon _{IJK}}{\text{tr}}({h_{{\alpha _{IJ}}}}[{h_{{e_K}}},\hat Q_v^{2n}]h_{{e_K}}^{ - 1})}  \hfill \\
   = 24(\hat H_E^{(n)}(v;e_x^ + ,e_y^ + ,e_z^ + ) + \hat H_E^{(n)}(v;e_x^ + ,e_y^ + ,e_z^ - )) = 48\hat \tilde H_E^{(n)}(v;e_x^ + ,e_y^ + ,e_z^ + ) \hfill \\
  \left( \begin{gathered}
  with:\hat H_E^{(n)}(v;e_x^ + ,e_y^ + ,e_z^ + ) + \hat H_E^{(n)}(v;e_x^ + ,e_y^ + ,e_z^ - ) = 2\hat \tilde H_E^{(n)}(v;e_x^ + ,e_y^ + ,e_z^ + ) \hfill \\
  and:[{h_{e_z^ + }},\hat Q_v^{2n}]h_{e_z^ + }^{ - 1} \to \sum\limits_{l{\text{is}}\;{\text{odd}}} {\sum\limits_{{\mathcal{P}_l}} {{{( - it\frac{{{{(\beta {a^2})}^3}}}{8})}^l}} } \hat Q_v^{{p_1}}{\varepsilon _{{\alpha _1}{\beta _1}{\gamma _1}}}{{\hat X}^{{\alpha _1}}}{{\hat Y}^{{\beta _1}}}{\tau ^{{\gamma _1}}} \hfill \\
   \times \hat Q_v^{{p_2}}{\varepsilon _{{\alpha _2}{\beta _2}{\gamma _2}}}{{\hat X}^{{\alpha _2}}}{{\hat Y}^{{\beta _2}}}{\tau ^{{\gamma _2}}} \cdots  \times \hat Q_v^{{p_l}}{\varepsilon _{{\alpha _l}{\beta _l}{\gamma _l}}}{{\hat X}^{{\alpha _l}}}{{\hat Y}^{{\beta _l}}}{\tau ^{{\gamma _l}}}\hat Q_v^{{p_{l + 1}}} \hfill \\ 
\end{gathered}  \right) \hfill \\ 
\end{gathered} 
\end{equation}
\end{figure*}

In fact, existing research results have long pointed out that the simplification of the Lorentz Hamiltonian is much more complicated than the simplification of the Euclidean Lorentzian. The current analytical calculation can only be derived step by step. For example, two order approximate expansion (note that the first term of the expansion is defined as zeroth order), you need to pass the commutation relationship $\displaystyle \frac{C}{t^{2}}\mathrm{tr}(h_{e_{x}^{+}}F_{1}h_{e_{x}^{+}}^{-1}h_{e_{y}^{+}}F_{2}h_{e_{y}^{+}}^{-1}G_{1})$. This involves calculating the expectation and eigenvalues of the first two levels of expansion of the field operator, and this can lead to the loss of some important expected values in the commutators of some terms (such terms are called special terms). Formula (2.14) can be further simplified to:

\begin{figure*}
\begin{equation}
\label{deqn_ex1a}
\begin{gathered}
  \hat H_L^{(\vec k)}(v) = \sum\limits_{\substack{{v_1},{v_2},{v_3},{v_4}, \\ {e_1},{e_J},{e_K}}}  {\hat H_L^{(\vec k)}}  {\hat H_L^{(\vec k)}} (v;{v_1},{v_2},{v_3},{v_4};{e_I},{e_J},{e_K}) \hfill \\
   = \sum\limits_{\substack{{v_1},{v_2},{v_3},{v_4}, \\ 
  {\text{  }}{e_1},{e_J},{e_K} }} {\frac{{ - 1}}{{2i{\beta ^7}{a^{10}}{t^5}}}{\varepsilon ^{IJK}}{\text{tr}}([{h_{{e_1}}},[\hat Q_{{v_1}}^{2{k_1}},\hat H_E^{({k_2})}({v_2})]] \times h_{{e_I}}^{ - 1}[{h_{{e_J}}},[\hat Q_{{v_3}}^{2{k_3}},\hat H_E^{({k_4})}({v_4})]]h_{{e_J}}^{ - 1}[{h_{{e_K}}},\hat Q_v^{2{k_5}}]h_{{e_K}}^{ - 1})}  \hfill \\
  {\text{ = }}48\sum\limits_{{v_1},{v_2},{v_3},{v_4}} {\hat \tilde H_L^{(\vec k)}} (v;{v_1},{v_2},{v_3},{v_4};e_x^ + ,e_y^ + ,e_z^ + ) \hfill \\
  \left( \begin{gathered}
  with:[\hat Q_{{v_1}}^{2k},\hat H_E^{(n)}({v_2};{e_l},{e_J},{e_K})] = \frac{2}{{i\beta {a^2}t}}({{\hat K}_1} + {{\hat K}_2}) \hfill \\ = \frac{2}{{i\beta {a^2}t}}\left( {\int {{{\text{d}}^3}} x\{ V,F_{ab}^i,(x)\} \frac{{{\varepsilon _{ijk}}(N({h_e}) \cdot E_i^a)(N({h_e}) \cdot E_j^b)}}{{\sqrt {\det (N({h_e}) \cdot E)} }}} \right) + \frac{2}{{i\beta {a^2}t}}{\varepsilon _{IJK}} \hfill \\ \sum\limits_{{p_1} + {p_2} = 2n - 1} {(2k)} \frac{{ - it{{(\beta {a^2})}^3}}}{8}{\text{tr}}(N({h_e}) \cdot {h_{dj}}{\tau ^\gamma })\hat Q_v^{{p_1}}\left[ {{{\hat Q}_v},{\varepsilon _{\alpha {\beta _1}\gamma }}(N({h_e}) \cdot \hat X_J^{{\alpha _1}})(N({h_e}) \cdot \hat X_J^{{\beta _1}})]\hat Q_v^{2k - 1 + {p_2}}} \right] \hfill \\
   + {\varepsilon _{IJK}}\sum\limits_{{p_1} + {p_2} = 2n - 1} {\frac{{2k(2k - 1)}}{2}} \frac{{ - it{{(\beta {a^2})}^3}}}{8}{\text{tr}}(N({h_e}) \cdot {h_{{\alpha _{IJ}}}}{\sigma ^\gamma })\hat Q_v^{{p_1}} \hfill \\ \left[ {{{\hat Q}_v},[{{\hat Q}_v},{\varepsilon _{a,{\beta _I}{\gamma _I}}}(N({h_e}) \cdot \hat X_I^{{\alpha _1}})(N({h_e}) \cdot \hat X_J^{{\beta _I}})]} \right] \hat Q_v^{2k - 2 + {p_2}} + O({t^4}) \hfill \\ 
\end{gathered}  \right)
\end{gathered} 
\end{equation}
\end{figure*}

\begin{equation}
\label{deqn_ex1a}
\begin{gathered}
  \hat H_L^{(\vec k)}(v;{v_1},{v_2},{v_3},{v_4},{e_l},{e_J},{e_K})
  { = ^{{\text{ext}}}}\hat H_L^{(\vec k)}(v;{v_1},{v_2},{v_3},{v_4};{e_l},{e_J},{e_K}) \hfill \\
  { + ^{{\text{alt}}}}\hat H_L^{(\vec k)}(v;{v_1},{v_2},{v_3},{v_4},{e_l},{e_J},{e_K}) \hfill \\ 
\end{gathered} 
\end{equation}

with, Of course, in the simplification of the above formula, it is inevitable to use the commutation relationship to rewrite the expected value as follows:

$$
\begin{gathered}
  {h_{{s^ + }}}\prod\limits_{i = 1}^m {(\sigma _i^ + \hat p_s^{{\alpha _i}}({e^ + }) + \sigma _i^ - \hat p_t^{{\alpha _i}}({e^ - }))} h_{{s^ + }}^{ - 1}
   \to \sum\limits_\mathcal{J} {{{(it)}^{|\mathcal{J}|}}} \prod\limits_{i \ne \mathcal{J}} {(\sigma _i^ + \hat p_s^{{\alpha _i}}({e^ + }) + \sigma _i^ - \hat p_t^{{\alpha _i}}({e^ - }))} \prod\limits_{j \in \mathcal{J}} {{\tau ^{{\alpha _j}}}}  \hfill \\ 
\end{gathered} 
$$

Finally, the following formula can be obtained. During the simplification process of this formula, the role of the field operator needs to be defined.

\begin{equation}
\label{deqn_exla}
\begin{gathered}
  \hat F(v;{v_1},{v_2},{v_3},{v_4};e_x^ + ,e_y^ + ,e_z^ + ): =  \hfill \\
  \hat H_L^{(\vec k)}(v;{v_1},{v_2},{v_3},{v_4};e_x^ + ,e_y^ + ,e_z^ + )
   + \hat H_L^{(\vec k)}(v;{\mathbf{r}}({v_3}),{\mathbf{r}}({v_4}),{\mathbf{r}}({v_1}),{\mathbf{r}}({v_2});e_x^ + ,e_y^ + ,e_z^ + ) \hfill \\ 
\end{gathered} 
\end{equation}

With:

$$
\begin{gathered}
 \begin{gathered}
  \hat F(v;{v_1},{v_2},{v_3},{v_4};e_x^ + ,e_y^ + ,e_z^ + )
   =  - {\text{tr}}([{h_{e_y^ + }},[\hat Q_{{\mathbf{r}}({v_1})}^{2{k_1}},\hat H_E^{({k_2})}({\mathbf{r}}({v_2}))]]h_{e_y^ + }^{ - 1} \hfill \\
  [{h_{e_x^ + }},[\hat Q_{{\mathbf{r}}({v_3})}^{2{k_3}},\hat H_E^{({k_4})}({\mathbf{r}}({v_4}))]] \cdot h_{e_x^ + }^{ - 1}[{h_{e_z^ + }},\hat Q_v^{2{k_5}}]h_{e_z^ + }^{ - 1}) \hfill \\
   + \operatorname{tr} ([{h_{e_x^ + }},[\hat Q_{{\mathbf{r}}({v_3})}^{2{k_1}},\hat H_E^{({k_2})}({\mathbf{r}}({v_4}))]]h_{e_x^ + }^{ - 1} \cdot 
  [{h_{e_y^ + }},[\hat Q_{{\mathbf{r}}({v_1})}^{2{k_3}},\hat H_E^{({k_4})}({\mathbf{r}}({v_2}))]]]h_{e_y^ + }^{ - 1}[{h_{e_z^ + }},\hat Q_v^{2{k_s}}]h_{e_z^ + }^{ - 1}) \hfill \\ 
\end{gathered} 
\end{gathered} 
$$

Expressing it explicitly, the final result can be obtained in equation(11). At the same time, the Lorentz term is showed in equation(2.12)

\begin{figure*}
\begin{equation}
\label{deqn_exla}
\begin{gathered}
 \begin{gathered}
  {\langle \widehat {{H_E}}\rangle _N} \approx {\mkern 1mu} 6a\left( {{N_0} + {{N'}_\eta }\eta  + \frac{1}{2}{{N''}_\eta }{\eta ^2} + \frac{1}{6}{{N'''}_\eta }{\eta ^3}} \right) \hfill \\
  \sqrt { - \beta \eta } \left( {{a_0} + {a_1}\cos (\xi ) + {b_1}\sin (\xi ) + {a_2}\cos (2\xi ) + {b_2}\sin (2\xi ) + {a_3}\cos (3\xi ) + {b_3}\sin (3\xi )} \right) \hfill \\
   - \frac{3}{4}a\left( {{N_0} + {{N'}_\eta }\eta  + \frac{1}{2}{{N''}_\eta }{\eta ^2} + \frac{1}{6}{{N'''}_\eta }{\eta ^3}} \right)t\sqrt { - \frac{\beta }{{{\eta ^3}}}} \left( {{a_0} + {a_1}\cos \left( {\frac{\xi }{2}} \right) + {b_1}\sin \left( {\frac{\xi }{2}} \right) +  \cdots } \right) \hfill \\
   \times \left( {{c_0} + {c_1}\cos \left( {\frac{\xi }{2}} \right) + {d_1}\sin \left( {\frac{\xi }{2}} \right) +  \cdots } \right) \hfill \\
   \times \left\{ {\cos \left( {\frac{\xi }{2}} \right)\left[ {8{\eta ^2} + 8\eta (4\cosh (\eta ) - 3){\text{cosh}}(\eta ) - 9} \right] - 12i\eta \sin \left( {\frac{\xi }{2}} \right)} \right\} \hfill \\
  n\left( {{\alpha _1}\eta \sinh (\eta ) + {\alpha _2}{\eta ^2}{{\cosh }^2}(\eta ) + {\alpha _3}{\eta ^3}\tanh (\eta )} \right)n \hfill \\
   \times \left( {{\beta _1}{{\cos }^2}\left( {\frac{\xi }{2}} \right) + {\beta _2}{{\sin }^2}\left( {\frac{\xi }{2}} \right) + {\beta _3}\sin \left( {\frac{\xi }{2}} \right)\cos \left( {\frac{\xi }{2}} \right)} \right)n + O({t^3})n \hfill \\ 
\end{gathered} 
\end{gathered} 
\end{equation}
\end{figure*}

\begin{figure*}
\begin{equation}
\label{deqn_exla}
\begin{gathered}
  \langle \widehat {{H_L}}\rangle  =  - \frac{{6a\left( {{N_0} + {{N'}_\eta }\eta  + \frac{1}{2}{{N''}_\eta }{\eta ^2} + \frac{1}{6}{{N'''}_\eta }{\eta ^3}} \right)\sqrt { - \beta \eta } }}{{{\beta ^2}}}{\left( {\sum\limits_{n = 0}^3 {{a_n}} \cos (n\xi ) + {b_n}\sin (n\xi )} \right)^2} \hfill \\
   - \frac{{3a\left( {{N_0} + {{N'}_t}t + \frac{1}{2}{{N''}_t}{t^2} + \frac{1}{6}{{N'''}_t}{t^3}} \right)t}}{{262144{{( - \beta \eta )}^{3/2}}}} \hfill \\
  \left\{ {2(3 - 220{\eta ^2})\left( {\sum\limits_{n = 0}^3 {{c_n}} \cos (6n\xi ) + {d_n}\sin (6n\xi )} \right)} \right. + 4i\eta \left( {\sum\limits_{n = 1}^5 {{e_n}} \sin (n\xi )} \right) 
  \hfill \\ + 2( - 3611 + 8\eta (492\eta  + 11i))\left( {\sum\limits_{n = 0}^3 {{f_n}} \cos (n\xi )} \right) \hfill \\- 2\left( { - 789 + 4\eta (305\eta  + 18i)} \right)
  \left( {\sum\limits_{n = 0}^3 {{g_n}} \cos (2n\xi )} \right) + \left( {4413 - 4\eta (928\eta  + 49i)} \right)\left( {\sum\limits_{n = 0}^3 {{h_n}} \cos (3n\xi )} \right) \hfill \\
   + 8( - 1978 + \eta (4192\eta  - 7i))\left( {\sum\limits_{n = 0}^3 {{j_n}} \cos (4n\xi )} \right) + ( - 7 + 4( - 272\eta  + 5i)\eta ) \hfill \\
  \left( {\sum\limits_{n = 0}^3 {{k_n}} \cos (5n\xi )} \right)\left. { - 4\eta \coth (\eta )\left[ {\sum\limits_{n = 0}^3 {{l_n}} \cos (n\xi )} \right] + 8\eta {\text{csch}}(\eta )\left[ {\sum\limits_{n = 0}^3 {{m_n}} \cos (n\xi )} \right]} \right\} \hfill \\
   + 8(1436 + \eta ( - 4056\eta  + 25i)) + \sum\limits_{n = 0}^3 {\left( {{p_n}\sinh (\eta )\cos (n\xi ) + {q_n}\cosh (\eta )\sin (n\xi )} \right)}  \hfill \\
  \sum\limits_{n = 0}^3 {\left( {{r_n}\tanh (\eta )\cos (2n\xi ) + {s_n}\coth (\eta )\sin (2n\xi )} \right)} 
   + \sum\limits_{n = 0}^3 {\left( {{u_n}{\text{csch}}(\eta )\cos (3n\xi ) + {v_n}{\text{sech}}(\eta )\sin (3n\xi )} \right)}  \hfill \\ + O({t^3}) \hfill \\ 
\end{gathered} 
\end{equation}
\end{figure*}

\subsection{Comparison of results between full model and simplified model}

First, a two-step simplification is performed on the above results: the hysteresis function is first simplified to 1, that is, there is no hysteresis effect in the Lorentz transformation. In addition, for the Euclidean model and the Lorentz model, only the ground state and the first excited state are calculated (that is, only two approximate expansions are performed), and the result obtained is:

\begin{equation}
\label{deqn_ex1a}
\begin{gathered}
\begin{gathered}
\begin{gathered}
  \langle \widehat {{H_E}}\rangle  = 6a\sqrt { - \beta \eta } {\sin ^2}(\xi ) - \frac{3}{4}at\sqrt { - \frac{\beta }{{{\eta ^3}}}} {\sin ^2}\left( {\frac{\xi }{2}} \right)\cos \left( {\frac{\xi }{2}} \right) \hfill \\
  \left\{ {\cos \left( {\frac{\xi }{2}} \right)[8{\eta ^2} + 8\eta (4\cosh (\eta ) - 3){\text{csch}}(\eta ) - 9]} \right. 
  \left. { - 12i\eta \sin \left( {\frac{\xi }{2}} \right)} \right\} + O({t^2}) \hfill \\ 
\end{gathered} 
\end{gathered} 
\end{gathered} 
\end{equation}

Similarly, note that $\langle\widehat{H_{L}}\rangle=\langle^{\mathrm{extr}}\widehat{H_{L}}\rangle+\langle^{\mathrm{alt}}\widehat{H_{L}}\rangle\rangle $, the approximate expansion expression of Lorentz's Hamiltonian can be written as equation(2.20)

\begin{figure*}
\begin{equation}
\label{deqn_ex1a}
\begin{gathered}
  \langle \widehat {{H_L}}\rangle  =  - \frac{{6\alpha \sqrt { - \beta \theta } {{\sin }^2}(\xi ){{\cos }^2}(\xi )}}{{{\rho ^2}}} - \frac{{3\alpha t}}{{262144{{( - 3\theta )}^{3/2}}}}[2(3 - 220{\eta ^2})\cos (6\xi ) + 4\eta (4838\sin (\xi ) \hfill \\ - 6284\sin (2\xi ) 
   + 4685\sin (3\xi ) - 5222\sin (4\xi ) - 108\sin (5\xi )) + 2( - 6611 + 8\eta (492\eta  + 11i)) \hfill \\ \cos (\xi ) - 2( - 2749 + 49(305 + 18i)
  \cos (2\xi ) + (441)3 - 4\eta (2829 + 49))\cos (2\xi ) + 8( - 8708  \hfill \\ + \eta (4192\eta  - 7i)) \cos (4\xi ) + ( - 7 + 4( - 272\eta  + 5i)\eta )\cos (5\xi ) 
   - 4\eta \coth (\eta )[56\cos (\xi ) + 1731\cos (2\xi )  \hfill \\ + 1524\cos (3\xi )- 40548\cos (4\xi ) + 116\cos (5\xi ) + 117\cos (6\xi ) + 57292]
   + 8\eta {\text{csch}}(\eta )[130\cos (\xi ) \hfill \\  + 918\cos (2\xi )+ 801\cos (3\xi ) - 18618\cos (4\xi ) + 125\cos (5\xi ) + 58\cos (6\xi ) + 16362] \hfill \\
   + 8[1436 + \eta ( - 4056\eta  + 25i)]) + O({t^2}) \hfill \\ 
\end{gathered}  
\end{equation}
\end{figure*}

From another perspective, if the hysteresis function is not simply taken as 1, but the sign function $N=\operatorname{sgn}(p_{b})\sqrt{|p_{a}|}$ (\cite{ref10}), because this allows analytical solutions for each conjugate dynamic variable. However, only the first three terms (that is, the first two level approximations) are taken, and the shtekar-Barbero variables and dynamic parameters are mapped according to the method in the article:

$$
\begin{gathered}
  A_1^1 =  - a,\quad A_2^2 =  - b,\quad A_3^3 =  - b\sin \theta ,\quad A_3^1 =  - \cos \theta  \hfill \\
  E_1^1 = |{p_a}|\sin \theta ,\quad E_2^2 = \frac{{{p_b}}}{2}\sin \theta ,\quad E_3^3 = \frac{{{p_b}}}{2} \hfill \\ 
\end{gathered} 
$$

with the exchange relationship: 

$$\displaystyle \{a,p_a\}=\frac{\kappa\beta}{8\pi},\quad\{b,p_b\}=\frac{\kappa\beta}{8\pi}
$$

and metric mapping relationship. This mapping relationship maps the Hilbert space complete basis in the previous theory to the selected dynamic coordinates.

$$
\begin{gathered}
  L_R^2{f^2} = \frac{{p_b^2}}{{4|{p_a}|}} = \frac{{{{({a^0})}^2}p_a^0}}{{{\beta ^2}}}4\sin {\left( {\frac{{T - {T_0}}}{2}} \right)^2}\cos {\left( {\frac{{T - {T_0}}}{2}} \right)^{ - 2}}
   = 4\frac{{{{({a^0})}^2}p_a^0}}{{{\beta ^2}}}\left( {\frac{{\sqrt {p_a^0} }}{r} - 1} \right) \hfill \\ 
\end{gathered} 
$$

Simply apply the corresponding mapping relationship, and you can write the first few terms of the Hamiltonian under this new hysteresis function and mapping relationship, as equation (2.21) shows.

\begin{figure*}
\begin{equation}
\label{deqn_ex1a}
\begin{gathered}
  H = \frac{{ - 4}}{{{\kappa ^2}\beta }}\sum\limits_{v \in V(\Gamma )} {\frac{1}{{{T_v}}}} \sum\limits_{i,j,k} \varepsilon  ({e_{v,i}},{e_{v,j}},{e_{v,k}})tr\left( {(h(\square _{ij}^v) - h{{(\square _{ij}^v)}^\dag })h{{({e_{v,k}})}^\dag }\{ h({e_{v,k}}),{V^\mu }\} } \right) \hfill \\
   + \frac{{64(1 + {\beta ^2})}}{{{\kappa ^4}{\beta ^7}}}\sum\limits_{v \in V(\Gamma )} {\frac{1}{{{T_v}}}} \sum\limits_{i,j,k} \varepsilon  ({e_{v,i}},{e_{v,j}},{e_{v,k}})tr\left( {h{{({e_{v,i}})}^\dag }\{ h({e_{v,i}}),K\} h{{({e_{v,j}})}^\dag }\{ h({e_{v,j}}),K\} } \right. \hfill \\
  \left. {h{{({e_{v,k}})}^\dag }\{ h({e_{v,k}}),{V^\mu }\} } \right) \hfill \\ 
\end{gathered} 
\end{equation}
\noindent\rule{\textwidth}{0.5pt}
\end{figure*}

with:

$$
\begin{gathered}
  {V^\mu }: = \sum\limits_{v \in {\text{V}}(\Gamma )} {\sqrt {\frac{1}{{{T_v}}}|{Q_v}|} } , \hfill \\
  {Q_v} = \sum\limits_{i,j,k} \varepsilon  ({e_{v,i}},{e_{v,j}},{e_{v,k}}){\varepsilon ^{IJK}}{E_I}({e_{v,i}}){E_J}({e_{v,j}}){E_K}({e_{v,k}}) \hfill \\ 
\end{gathered} 
$$

The semiclassical expansion expressions of the first two terms are shown below.

\begin{figure*}
\begin{equation}
\label{deqn_ex1a}
\begin{gathered}
  H =  - N\frac{{8\pi {\text{sgn}}({p_b})}}{{\kappa {\beta ^2}\sqrt {|{p_a}|} }}\left[ {2ab|{p_a}| + ({b^2} + {\beta ^2})2ab|{p_a}| + ({b^2} + {\beta ^2})\frac{{{p_b}}}{2}} \right. - \frac{{1 + {\beta ^2}}}{{{\beta ^2}}}b(\cos ({\mu _a}a) \hfill \\
   + \cos ({\mu _b}b))\left. {\left( {|{p_a}|a\cos ({\mu _b}b) + \frac{{{p_b}}}{8}b(\cos ({\mu _a}a) + \cos ({\mu _b}b))} \right)} \right] + N\frac{{\pi {\text{sgn}}({p_b})}}{{144\kappa {\beta ^2}\sqrt {|{p_a}|} }} \hfill \\
  [96\mu _1^2{a^2}b\left( {2a\left( {3{\beta ^2} + 5} \right)|{p_a}|} \right.\left. { + 3b\left( {{\beta ^2} + 1} \right){p_b}} \right) + 24\mu _2^2\left( {2ab|{p_a}|\left( {18{b^2}{\beta ^2} + 2\left( {11{b^2} + 6{\beta ^2}} \right) + 17} \right)} \right. \hfill \\
  \left. { + {p_b}\left( {2\left( {3{\beta ^2} + 5} \right){b^4} + 3{b^2} + 7{\beta ^2}} \right)} \right)n + \mu _3^2\left( {2ab|{p_a}|\left( {288{b^2}{\beta ^2} + \left( {352{b^2} + 3{\beta ^2}} \right) + 59} \right)} \right. \hfill \\
  \left. {\left. { + {b^2}{p_b}\left( {96{b^2}{\beta ^2} + 5\left( {32{b^2} + 9{\beta ^2}} \right) - 19} \right)} \right)} \right] \hfill \\ 
\end{gathered} 
\end{equation}
\noindent\rule{\textwidth}{0.5pt}
\end{figure*}

First, peel off its 0th-order, 1st-order and 2nd-order expressions. Note that in order to contrast with the existing research results, we adopt the same quantization scheme:
$\displaystyle a\to\frac{\sin(\mu_aa)}{\mu_a},\quad b\to\frac{\sin(\mu_bb)}{\mu_b}$. The result after the first-order expansion is:

\begin{equation}
\label{deqn_ex1a}
\begin{gathered}
 {H_{cl}} =  - N\frac{{8\pi }}{{\kappa {\beta ^2}}}\frac{{\operatorname{sgn} ({p_b})}}{{\sqrt {\left| {{p_a}} \right|} }}\left[ {2ab\left| {{p_a}} \right| + \left( {{b^2} + {\beta ^2}} \right)\frac{{{p_b}}}{2}} \right]
\end{gathered} 
\end{equation}

Among them, $a, b, p_a, p_b$ are four dynamic parameters obtained according to Ashtekar-Barbero formulation of GR \cite{ref22,ref23,ref24}, $a, b$ describe the behavior of space (such as scaling, etc.), $p_a, p_b$ describes its conjugate momentum $\left| {{p_a}} \right| = L_S^2{g^2},\left| {{p_b}} \right| = 2{L_R}{L_S}\left| {fg} \right|$. Among the prefactors, $\kappa = 16\pi G/c^3$ includes the gravitational constant and the speed of light. The lag function $N$ describes the change of the time coordinate under the general theory of relativity. The second term describes the change of the space geometry, The dynamic term of the Hamiltonian corresponds to the energy change caused by the spatial geometric transformation. $\beta$ is the semi-classical Barbero-Immirzi parameter, used to describe the behavior at extreme points. Take advantage of transformations: $\displaystyle a\to\frac{\sin(\mu_aa)}{\mu_a},\quad b\to\frac{\sin(\mu_bb)}{\mu_b}$ The corresponding quantized model can be obtained:

\begin{equation}
\label{deqn_ex1a}
\begin{gathered}
  H_{{\text{eff}}}^{(1)} =  - N\frac{{8\pi }}{{\kappa {\beta ^2}}}\frac{{{\text{sgn}}({p_b})}}{{\sqrt {|{p_a}|} }}\left[ {2\frac{{\sin ({\mu _a}a)}}{{{\mu _a}}}\frac{{\sin ({\mu _b}b)}}{{{\mu _b}}}|{p_a}|} \right. \left. {\left( {\frac{{\sin {{({\mu _b}b)}^2}}}{{\mu _b^2}} + {\beta ^2}} \right)\frac{{{p_b}}}{2}} \right] \hfill \\ 
\end{gathered} 
\end{equation}

The discrete nature of sinusoidal quantities determines this quantum property. where the sine parameters are determined by the metric of the spatial coordinates:$ds^{2}=-N(T)^{2}dT^{2}+L_{R}^{2}f(T)^{2}dR^{2}+L_{S}^{2}g(T)^{2}d\Omega^{2}$

Regularizing the classical Hamiltonian according to the LQC theory, we can get \cite{ref25}:

\begin{equation}
\label{deqn_ex1a}
\begin{gathered}
  H_{{\text{eff}}}^{(2)} = N\frac{{8\pi {\text{sgn}}({p_b})}}{{\kappa \sqrt {|{p_a}|} }}\left[ {2\frac{{\sin ({\mu _a}a)}}{{{\mu _a}}}\frac{{\sin ({\mu _b}b)}}{{{\mu _b}}}|{p_a}| + } \right. \hfill \\
  {\text{       }}\left( {\frac{{\sin {{({\mu _b}b)}^2}}}{{\mu _b^2}} - 1} \right)\frac{{{p_b}}}{2} - \frac{{1 + {\beta ^2}}}{{{\beta ^2}}}\frac{{\sin ({\mu _b}b)}}{{{\mu _b}}}cos(\mu_i a_i) \hfill \\
  {\text{       }}\left. {\left( {|{p_a}|\frac{{\sin ({\mu _a}a)}}{{{\mu _a}}}\cos ({\mu _b}b) + \frac{{{p_b}}}{8}\frac{{\sin ({\mu _b}b)}}{{{\mu _b}}}cos(\mu_i a_i)} \right)} \right] \hfill \\ 
\end{gathered} 
\end{equation}

Where $cos(\mu_i a_i)=cos(\mu_a a)+cos(\mu_b b)$. 
Mathematically speaking, the difference between these two results is due only to the different regularization mappings used: 

$$\begin{gathered}
  {\iota _{A,E}}[K_x^I(y)] =  - \frac{4}{{\kappa {\beta ^3}{\mu _x}}}
  {\text{Tr}}\left[ {{\tau ^I}h{{({e_{y,x}})}^\dag }\{ h({e_{y,x}}),\{ {\iota _{A,E}}[\int_\sigma  {\text{d}} {x^3}{H_E}(x)],{\iota _{A,E}}[V]\} \} } \right] \hfill \\ 
\end{gathered} $$

and 

$$\begin{gathered}
  {\iota _2}[K_x^I(y)] =  - \frac{4}{{\kappa {\beta ^3}{\mu _x}}}
  {\text{Tr}}\left[ {{\tau ^I}\tilde h{{({e_{y,x}})}^\dag }\{ \tilde h({e_{y,x}}),\{ \int_\sigma  {\text{d}} {x^3}{\iota _1}[\omega ({H_E}(x))],{\iota _1}[\omega (V)]\} \} } \right] \hfill \\ 
\end{gathered} $$

This is because the order of symplectic reduction and regularization is inconsistent and cannot be changed at will. The latter method is based on the method of first symplectic simplification and then regularization, but it refers to the idea of partial complete theory and takes into account partial quantum effects and corrections. It is not difficult to see that this result is also determined by the second-level expansion of equation (2.22).
Finally, consider the last level of quantum correction and use the $\mu_0$ method to perform the following variable mapping:

$$
\begin{gathered}
  {\mu _1} = \frac{1}{{{N_1}}},\quad {\mu _2} = \frac{\pi }{{{N_2} + 1}},\quad {\mu _3} = \frac{{2\pi }}{{{N_3}}} \hfill \\
  {R_v} \in \left\{ {{\mu _1},2{\mu _1},...,1} \right\},\quad {\theta _v} \in \left\{ {{\mu _2},2{\mu _2},...,\pi  - {\mu _2}} \right\}, \quad {\varphi _v} \in \left\{ {{\mu _3},2{\mu _3},...,2\pi } \right\} \hfill \\ 
\end{gathered} 
$$

The third level expansion is (using the relation $\displaystyle H=\int_\sigma\mathrm{d}x^3N(x)(H_E(x)+H_L(x))$)

\begin{equation}
\label{deqn_ex1a}
\begin{gathered}
  H_{{\text{eff}}}^{(3)}: =  - N\frac{{8\pi {\text{sgn}}({p_b})}}{{\kappa {\beta ^2}\sqrt {|{p_a}|} }}\left[ {2ab|{p_a}| + ({b^2} + {\beta ^2})\frac{{{p_b}}}{2}} \right] \hfill \\
   + N\frac{{\pi {\text{sgn}}({p_b})}}{{144\kappa {\beta ^2}\sqrt {|{p_a}|} }}[96\mu _1^2{a^2}b(2a\left( {3{\beta ^2} + 5} \right)|{p_a}| + 3b\left( {{\beta ^2} + 1} \right){p_b}) \hfill \\
   + 24\mu _2^2(2ab|{p_a}|\left( {18{b^2}{\beta ^2} + 2\left( {11{b^2} + 6{\beta ^2}} \right) + 17} \right)
   + {p_b}\left( {2\left( {3{\beta ^2} + 5} \right){b^4} + 3{b^2} + 7{\beta ^2}} \right)) \hfill \\
   + \mu _3^2\left( {2ab|{p_a}|\left( {288{b^2}{\beta ^2} + \left( {352{b^2} + 3{\beta ^2}} \right) + 59} \right)} \right.
  \left. { + {b^2}{p_b}\left( {96{b^2}{\beta ^2} + 5\left( {32{b^2} + 9{\beta ^2}} \right) - 19} \right)} \right)] \hfill \\ 
\end{gathered} 
\end{equation}

It can be obtained from the previous third-order expansion of the theoretical general formula. Except for the difference of a second-order small quantity of the quantum correction quantity, it is consistent with the expression provided by \cite{ref10}. Therefore, the general formula of the third-order theory we calculated in this article further expands the LQG model and is in good agreement with existing research results. For the sake of intuition, we reproduce and expand the numerical calculation results after assigning this result in Figure 1. The numerical calculation results and theoretical results obtained by different methods and simplified models are plotted in a picture. From the picture, we can see the degree of agreement and difference between the Hamiltonian simulated cosmological models under different models. It is easy to see from this image the difference in numerical solution and evolution of the three-level Hamiltonian expansion (corresponding to two different LQG models and a more "analytic" model). It can be found that singular points appear starting from the second-level expansion, which well expresses the nature of the extreme positions of the cosmological black hole model.

\begin{figure*}[!t]
\centering
\includegraphics[width=6in]{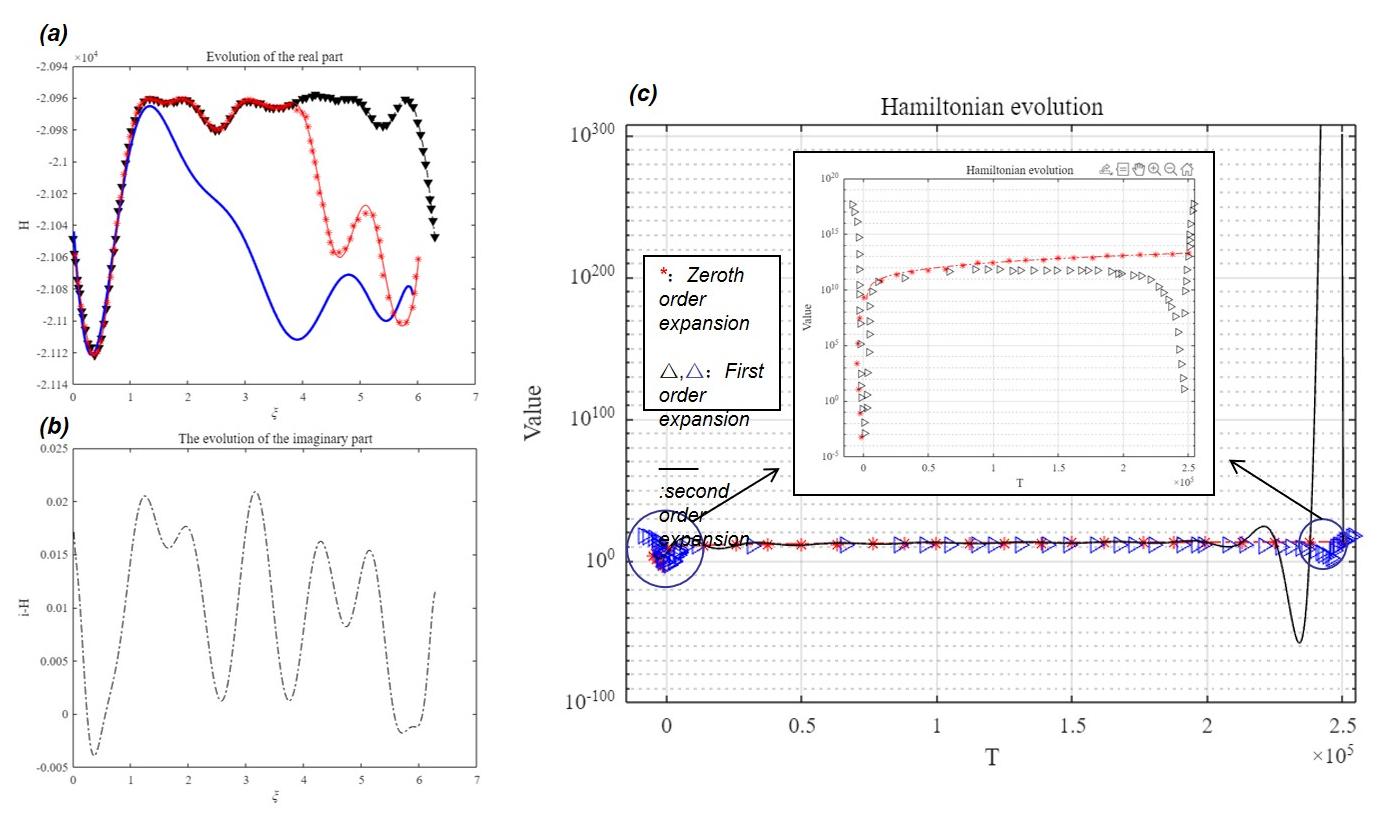}
\caption{(a) The third-level expansion of the real part of $H_L$. (b) The expansion part of the imaginary number. (c) The evolution of the third-level Hamiltonian. The circled part expresses the singular points that appear in the evolution process, and from the second-level The expansion begins to appear.}
\label{fig_4}
\end{figure*}

The work of this paper has completed the expansion and computational verification of the analytical model, and the following preliminary conclusions can be drawn: To achieve a reasonable description of the dynamics of black holes (including the propagation of quantum information), at least a second-level expansion is required, which embodies the LQG model. Compared with the advantages of the traditional Hamiltonian model, it is currently not possible to accurately describe the differences in quantum information transmission and expansion above the second level.

\section{Encoding and decoding quantum information in black holes}
\subsection{Decoding quantum information in black holes}

In this section, we will use the LQG model introduced previously to explore the encoding and decoding mechanism of quantum information. But before that we should note that microscopic models were generally used to describe black holes with extreme properties in the past \cite{ref31}. This microscopic model divides the interior of the black hole into many small spaces, called "two-dimensional anti-de Sitter space (AdS2)". Although these throats have a certain symmetry in theory (called $SL(2, R)$ symmetry ), but this symmetry may be broken due to interactions between larynxes. Although parts of the symmetry are broken, an overall symmetry remains. This means that some basic properties of black holes remain unchanged. In the original structure of the black hole, there are some low-energy states or modes. These patterns are related to the black hole's breakup process. Each throat has a certain mass, which makes them play an important role in the physics of the black hole.

Researchers in \cite{ref32} proposed a very simplified model, which is based on the fact that the splitting process of gravitational instantons is difficult to solve using the controlled perturbation method \cite{ref33,ref34}, and simplified the model into a quantum oscillator. However, looking at the solution results, we found that this model is accurate enough for large charges, etc., but unfortunately it still cannot explain the quantum information paradox problems such as space singularities. Therefore, this article directly starts from the H third-order approximate analytical formula in section II to solve the wave function. Assume that a semi-classical information particle passes through a black hole with energy $E$, and the black hole takes the effective potential energy $H_{eff}^{(1)},H_{eff}^{(2)},H_{eff}^{(2)}$. For the sake of simplicity, below we only express it as a symbol $H$.

According to the Schrödinger equation $\displaystyle \frac{d^{2}\psi}{dx^{2}}=-\frac{p^{2}}{\hbar^{2}}\psi $ with $\displaystyle p\left(x\right)\equiv\sqrt{2m\left[E-H \right]}$. Assume that the wave function still has the following vibration mode: $\displaystyle \psi (A_a^b,E_c^d) = A(A_a^b,E_c^d){e^{i\phi (A_a^b,E_c^d)}}$. Assume that the wave function still has the following vibration mode: $\displaystyle \psi(x_i)=A(x_i)e^{i\phi(x_i)}$. The result after derivation is as follows. Note that the symbolic expression $x_i$ represents one of the generalized coordinates.

$$
\begin{gathered}
  \frac{{\partial \psi }}{{\partial {x_i}}} = \left( {{A^\prime } + iA{\phi ^\prime }} \right){e^{i\phi }} \hfill \\
  {\nabla ^2}\psi  = [A'' + 2iA{\text{'}}\phi {\text{'}} + iA\phi '' - A{\left( {\phi ''} \right)^2}]{e^{i\phi }} \hfill \\ 
\end{gathered} 
$$

From this we get:

$$
A^{''}+2iA^{'}\phi^{'}+iA\phi^{°}-A\left(\phi^{\prime}\right)^{2}=-\frac{p^{2}}{\hbar^{2}}A
$$

Using the idea of slow-varying approximation, assuming that the change in amplitude is slowly changing, the wave function has the following approximate form:

$$
\displaystyle \psi  \cong \frac{C}{{\sqrt p }}{e^{ \pm \frac{i}{h}\iiint {pdxdydz}}}
$$

Consider the tunneling effect in a black hole, where $p$ is an imaginary number. Therefore the solution to the wave function in the region where quantum effects are taken into account (the tunneling region) may follow the form:

$$
\begin{gathered}
  \psi \left( {A,E} \right) \cong \frac{C}{{\sqrt {\left| {p(A,E)} \right|} }}{e^{\frac{1}{\hbar }{{\iiint {\left| {p\left( {A,E} \right)} \right|dAdE}}^ \cdot }}}  + \frac{D}{{\sqrt {\left| {p(A,E)} \right|} }}{e^{ - \frac{1}{\hbar }{{\iiint {\left| {p\left( {A,E} \right)} \right|dAdE}}^ \cdot }}} \hfill \\ 
\end{gathered} 
$$

Similar to WKB's approximation theory, although the form of this wave function can be considered to be true individually, it may not be true at the connection. This is a major limitation of this approximation method. Therefore, in the case of the above hypothetical solution, it is necessary to test whether these semi-classical regions and quantum regions still satisfy the connection relationship just like the traditional potential field. Assume that its general solution still has the form:

\begin{equation}
\label{deqn_ex1a}
\psi = \left\{ \begin{gathered}
  \frac{1}{{\sqrt {p(A,E)} }} \cdot {\text{     (region 1)}} \hfill \\
  \left[ {B{e^{\frac{i}{\hbar }\iiint {p(A,E)dAdE}}} + C{e^{ - \frac{i}{\hbar }\iiint {p(A,E)dAdE}}}} \right] \hfill \\
  \frac{1}{{\sqrt {p(A,E)} }}D{e^{\frac{i}{\hbar }\iiint {\left| {p(A,E)} \right|p(A,E)dAdE}}}({\text{region 2}}) \hfill \\ 
\end{gathered}  \right.
\end{equation}

Expand the potential energy at the junction:

$$
\begin{gathered}
  H(A,E) \approx H({A_0},{E_0}) + {\left. {\frac{{\partial H}}{{\partial A}}} \right|_{({A_0},{E_0})}}(A - {A_0}) + \hfill \\
  {\left. {\frac{{\partial H}}{{\partial E}}} \right|_{({A_0},{E_0})}}(E - {E_0}) + \frac{1}{2}{\left. {\frac{{{\partial ^2}H}}{{\partial {A^2}}}} \right|_{({A_0},{E_0})}}{(A - {A_0})^2} + \frac{1}{2} \cdot  \hfill \\
  {\left. {\frac{{{\partial ^2}H}}{{\partial {E^2}}}} \right|_{({A_0},{E_0})}}{(E - {E_0})^2} + \left. {\frac{{{\partial ^2}H}}{{\partial A\partial E}}} \right|(A - {A_0})(E - {E_0}) \cdots  \hfill \\ 
\end{gathered} 
$$

By linearly expanding the potential energy we get:

$$
\begin{gathered}
  H({A_0},{E_0})\psi  + {\left. {\frac{{\partial H}}{{\partial A}}} \right|_{({A_0},{E_0})}}(A - {A_0})\psi  \hfill \\
   + {\left. {\frac{{\partial H}}{{\partial E}}} \right|_{({A_0},{E_0})}}(E - {E_0})\psi  + \frac{1}{2}{\left. {\frac{{{\partial ^2}H}}{{\partial {A^2}}}} \right|_{({A_0},{E_0})}}{(A - {A_0})^2}\psi  \hfill \\
   + \frac{1}{2}{\left. {\frac{{{\partial ^2}H}}{{\partial {E^2}}}} \right|_{({A_0},{E_0})}}{(E - {E_0})^2}\psi 
   + {\left. {\frac{{{\partial ^2}H}}{{\partial A\partial E}}} \right|_{({A_0},{E_0})}}(A - {A_0})(E - {E_0})\psi  = E\psi  \hfill \\ 
\end{gathered} 
$$

The expansion formulas of each order can be obtained:

$$
\left\{ {\begin{array}{*{20}{l}}
  {H({A_0},{E_0})\psi  = {E_0}\psi ,} \\ 
  {{{\left. {\frac{{\partial H}}{{\partial A}}} \right|}_{({A_0},{E_0})}}(A - {A_0})\psi  + {{\left. {\frac{{\partial H}}{{\partial E}}} \right|}_{({A_0},{E_0})}}(E - {E_0})\psi  = {E_1}\psi ,} \\ 
  \begin{gathered}
  \frac{1}{2}{\left. {\frac{{{\partial ^2}H}}{{\partial {A^2}}}} \right|_{({A_0},{E_0})}}{(A - {A_0})^2}\psi  + \frac{1}{2}{\left. {\frac{{{\partial ^2}H}}{{\partial {E^2}}}} \right|_{({A_0},{E_0})}}{(E - {E_0})^2}\psi  \hfill \\
   + {\left. {\frac{{{\partial ^2}H}}{{\partial A\partial E}}} \right|_{({A_0},{E_0})}}(A - {A_0})(E - {E_0})\psi  = {E_2}\psi  \hfill \\ 
\end{gathered}
\end{array}} \right.
$$

The first two orders of expansion can use the Airy equation $\displaystyle \frac{d^2\psi_p}{dz^2}=z\psi_p$ Write down its analytical order, but starting from the third order expansion, the current theory can only obtain the results of numerical solutions through analytical methods for the time being. For the time being, we will not discuss how much impact the three-level expansion will have on the final result. We will only consider the approximate expansion of the first two levels for the time being.

Let: $I = p(x) \cong \sqrt {2m\left( {E - ({H_0} + \frac{{\partial H}}{{\partial A}}(A - {A_0})...)} \right)} $ Then the wave function and patched wave function can be obtained:

$$
{\psi _p}\left( x \right) \cong \frac{a}{{3\sqrt \pi  {I^{1/6}}}}{e^{ - I}} + \frac{{2b}}{{3\sqrt \pi  {I^{1/6}}}}{e^I}
$$

The connection conditions can be obtained by comparing the patch function and the wave function. The results of numerical verification confirm that in the given third-order expansion effective Hamiltonian case, the approximate analytical solution is valid. In subsequent calculations, $\psi$ and $\ E_n$ represents the wave function and energy level. Next, its time evolution is studied. Suppose there is a momentum invariant operator $\hat{I}$ \cite{ref35,ref36}. 

\begin{equation}
\label{deqn_ex1a}
\frac{d\hat{I}}{du}\equiv\frac{\partial\hat{I}}{\partial u}+\frac{i}{\hbar}[\hat{H},\hat{I}]=0.
\end{equation}

Evolving quantum states in the Schrödinger picture: $\hat{I}\left|\psi_{n}\right\rangle=c_{n}\left|\psi_{n}\right\rangle$
Then the evolution equation is:

\begin{equation}
\label{deqn_ex1a}
-\frac{i}{\hbar}\hat{H}\hat{I}\left|\psi_{n}\right\rangle+\frac{i}{\hbar}\hat{I}\hat{H}\left|\psi_{n}\right\rangle+\hat{I}\left|\psi_{n}\right\rangle=c_{n}|\psi_{n}\rangle+\dot{c_{n}}\left|\psi_{n}\right\rangle 
\end{equation}

Along with the initial conditions: $0=\dot{c_n}.$ Bringing the explicit Hamiltonian into the evolution equation gives equation(3.4):

\begin{figure*}
\begin{equation}
\label{deqn_ex1a}
\begin{gathered}
   - \sum {\frac{{\partial \hat I}}{{\partial {u_i}}}}  = \left( { - N\frac{{8\pi {\text{sgn}}({p_b})}}{{\kappa {\beta ^2}\sqrt {|{p_a}|} }}\left[ {2ab|{p_a}| + ({b^2} + {\beta ^2})\frac{{{p_b}}}{2}} \right] + N\frac{{\pi {\text{sgn}}({p_b})}}{{144\kappa {\beta ^2}\sqrt {|{p_a}|} }}} \right. \hfill \\
  \left[ {96\mu _1^2{a^2}b\left( {2a(3{\beta ^2} + 5)|{p_a}| + 3b({\beta ^2} + 1){p_b}} \right) + 24\mu _2^2\left( {2ab|{p_a}|(18{b^2}{\beta ^2} + 2(11{b^2} + 6{\beta ^2}) + 17)} \right.} \right. \hfill \\
  \left. { + {p_b}(2(3{\beta ^2} + 5){b^4} + 3{b^2} + 7{\beta ^2})} \right) + \mu _3^2\left( {2ab|{p_a}|(288{b^2}{\beta ^2} + (352{b^2} + 3{\beta ^2}) + 59) + {b^2}{p_b}} \right. \hfill \\
  \left. {\left. {\left( {(96{b^2}{\beta ^2} + 5(32{b^2} + 9{\beta ^2}) - 19)} \right)} \right)} \right]\hat I - \hat I\left( { - N\frac{{8\pi {\text{sgn}}({p_b})}}{{\kappa {\beta ^2}\sqrt {|{p_a}|} }}\left[ {2ab|{p_a}| + ({b^2} + {\beta ^2})\frac{{{p_b}}}{2}} \right]} \right. \hfill \\
  \left( { + N\frac{{\pi {\text{sgn}}({p_b})}}{{144\kappa {\beta ^2}\sqrt {|{p_a}|} }}\left[ {96\mu _1^2{a^2}b\left( {2a(3{\beta ^2} + 5)|{p_a}| + 3b({\beta ^2} + 1){p_b}} \right)} \right.} \right. + 24\mu _2^2 \hfill \\
  \left( {2ab|{p_a}|(18{b^2}{\beta ^2} + 2(11{b^2} + 6{\beta ^2}) + 17) + {p_b}(2(3{\beta ^2} + 5){b^4} + 3{b^2} + 7{\beta ^2})} \right) \hfill \\
  \left. {\left. { + \mu _3^2\left( {2ab|{p_a}|(288{b^2}{\beta ^2} + (352{b^2} + 3{\beta ^2}) + 59) + {b^2}{p_b}(96{b^2}{\beta ^2} + 5(32{b^2} + 9{\beta ^2}) - 19)} \right)} \right]} \right) \hfill \\ 
\end{gathered} 
\end{equation}
\noindent\rule{\textwidth}{0.5pt}
\end{figure*}

The expression of the momentum invariant operator is approximately:

\begin{figure*}[!t]
\centering
\includegraphics[width=6in]{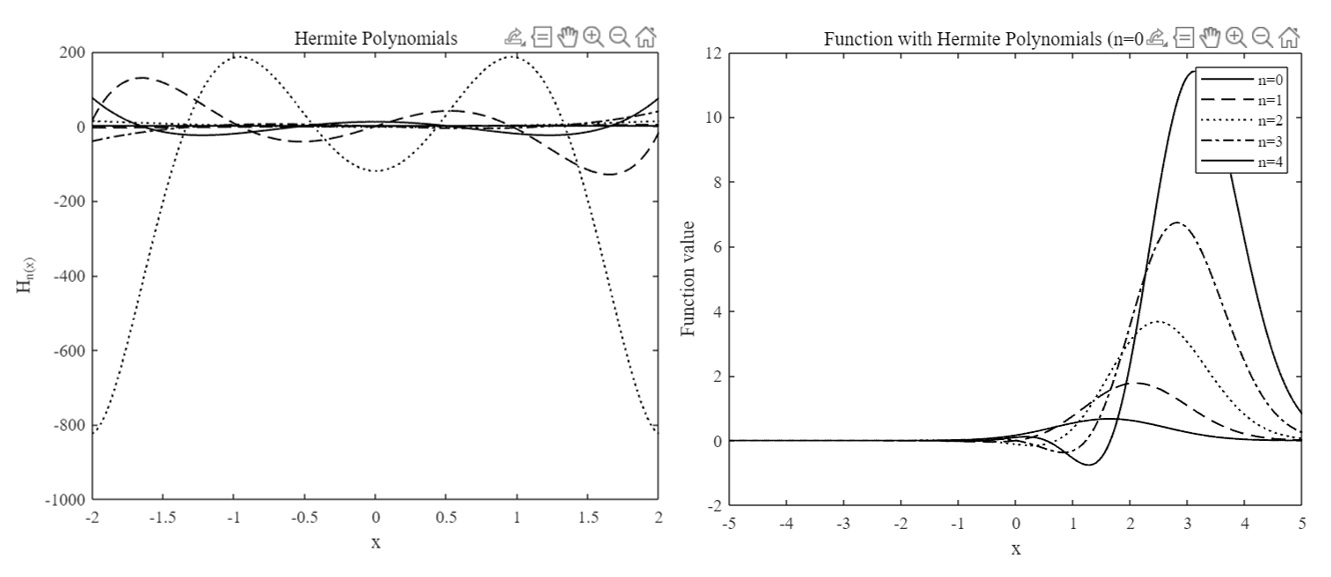}
\caption{Numerical solution results of Hermite polynomials$\begin{aligned}H_n(x)=2xH_{n-1}(x)-2(n-1)H_{n-2}(x)\end{aligned}$ and  Numerical solution results of $F_n$.$\displaystyle 
\left(\frac{\omega}{\pi}\right)^{1/4}\frac{1}{\sqrt{2^{n}n!}}H_{n}\left[\sqrt {\frac{\omega}{h}}(x-x_{p}(u))\right] \times\exp\left(i\frac{p_{p}(u)}{\hbar}x- \frac{\omega}{2\hbar}(x-x_{p}(u))^{2}\right)\exp(-iF_{n}(u))
$ (take quantum resonator as an example)}
\label{fig_4}
\end{figure*}

\begin{equation}
\label{deqn_ex1a}
\begin{gathered}
  \hat I = \frac{1}{2}\left[ {A\left( {p - {p_a}} \right) + A\left( {p - {p_b}} \right) + B{{\left( {p - {p_a}} \right)}^2} + } \right. 
  \left. {B{{\left( {p - {p_b}} \right)}^2} + C\left( {p - {p_a}} \right)\left( {p - {p_b}} \right) +  \cdots } \right] \hfill \\
   + F\left( {A_i^j,E_i^j,{\mu _1},{\mu _2},{\mu _3} \ldots  \ldots } \right) \cdot  
  \left[ {D\left( {{A_1} - a} \right) + D\left( {{E_1} - b} \right) + E{{\left( {{A_1} - a} \right)}^2} + E{{\left( {{E_1} - b} \right)}^2}} \right. \hfill \\
  \left. { + F\left( {{A_1} - a} \right)\left( {{E_1} - b} \right) \cdots } \right] \hfill \\ 
\end{gathered} 
\end{equation}

The next solution process should be to find the eigenvalue of the operator $\hat{I}$. However, this step cannot yet yield an analytical solution based on the current theory. All we can know is that for the first and second order expansions, it can be similar to the eigenvalues of the quantum oscillator, expressed in the following form:

\begin{equation}
\label{deqn_ex1a}
\begin{gathered}
  \langle x|{\phi _n}(A,E;a,b,{p_a},{p_b})\rangle  = F'\left( {A,E,{\mu _1},{\mu _2},{\mu _3}} \right) 
  {H_n}\left[ {F''\left( {A,E,{\mu _1},{\mu _2},{\mu _3}} \right)({A_1} - a)({E_1} - b)} \right] \hfill \\
   \times \exp \left( {\sum {i\frac{{{p_a}}}{\hbar }a}  - \frac{{F'''\left( {A,E,{\mu _1},{\mu _2},{\mu _3}} \right)}}{{2\hbar }}} \right. \hfill \\
  \left. {\left[ {{{({A_1} - a)}^2} + {{({E_1} - b)}^2} + ({A_1} - a)({E_1} - b)} \right]} \right)\exp ( - i{F_n}) \hfill \\ 
\end{gathered}
\end{equation}

\text{where } $H_n$ \text{ is the } $n-th$ order Hermite polynomial:

$$
{F_n} = \hat I + h\iiint {{\text{d}}a{\text{d}}b{\text{d}}{p_a}{\text{d}}{p_b}}\prod {G(a,b,{p_a},{p_b})} 
$$

Where $G_n$ represents the differential equation (s) regarding variables ($a,b,p_a,p_b$) determined by equation (30). For equations expanded to third order and above, numerical solution methods need to be used. We will put this part of the work in section IV. 

This theory states the relationship between quantum states and their corresponding classical phase space trajectories. However, the prerequisite for these states to be called eigenstates is that they cannot be driven by external fields. The excited state corresponding to the coherent state is called the coherent excited state, which is a set of bases, and any state can be expressed as a linear combination of this set of bases.

Therefore, the following numerical calculation algorithm is about solving the system of differential equations shown in equations (30) and the equation about the mass of the black hole: 

\begin{equation}
\label{deqn_ex1a}
\begin{gathered}
  \dot M(u) =
  \frac{{16\pi G}}{{{\Phi _r}}}\gamma \left( {\Delta O(u)\langle \dot \hat x(u)\rangle  + (\Delta  - 1)\dot O(u)\langle \hat x(u)\rangle } \right) \hfill \\ 
\end{gathered} 
\end{equation}

and the Klein-Gordon equation.

\begin{equation}
\label{deqn_ex1a}
\partial_{r}\left(\partial_{u}\chi-\frac12\left(1-r^{2}M(u)\right)\partial_{r}\chi\right)+\frac{\Delta(\Delta-1)}{2r^{2}}\chi=0 
\end{equation}

With \cite{ref37}:

\begin{equation}
\label{deqn_ex1a}
\begin{gathered}
  \chi (r,u) \approx {\chi ^{(0)}}(u){r^{1 - \Delta }} + {\chi ^{(0)}}(u){r^{2 - \Delta }} + O(u){r^\Delta } +  \ldots  \hfill \\
  \Phi  = \frac{{{\Phi _r}}}{r} + 16\pi G\left( {\frac{{{\chi ^{(0)}}{{(u)}^2}}}{4}{r^{ - 1/2}} + \int_0^r d {r^{\prime \prime }}\frac{1}{{{r^{\prime \prime 2}}}}} \right. \hfill \\
  \left. {\left( {\frac{1}{8}{\chi ^{(0)}}{{(u)}^2}r'{'^{\frac{1}{2}}} - \int_0^{{r^{\prime \prime }}} d {r^\prime }{r^{\prime 2}}{{\left( {{\partial _{{r^\prime }}}\chi ({r^\prime },u)} \right)}^2}} \right)} \right) \hfill \\ 
\end{gathered} 
\end{equation}

We devote some space to explaining how this evolution equation can be used to determine the mechanism by which quantum information is encoded and decoded in a black hole. Under our current theoretical framework, the process of information-carrying examples propagating information to a black hole is approximated as the energy exchange between semi-classical particles (such as quantum resonators) and the black hole. Therefore, we have placed in the appendix a solution that is also applicable to the WKB approximation method The process of quantum oscillator equation. Think of this process as a classical or semi-classical one. If the evolution approaches this classical solution before and after entering and leaving the black hole, we consider that the encoding and decoupling of information is complete. A decoupled quantum oscillator follows a quantum trajectory that completely, but non-isometrically, encodes its initial state. Information about the oscillator's initial state is also encoded into the black hole's late oscillations, and over time the black hole's mass eventually stabilizes. Therefore, this is a temporary copy of information that can be encoded into the degrees of freedom of Hawking radiation and the black hole's "hair".

One of the characterization schemes is the incomplete transfer of energy to the black hole through quantum resonators and the mass evolution equation of the black hole.

$$E_{\mathrm{tot}}=E_{\mathrm{osc}}+M_{\mathrm{ADM}}$$

Where:

\[\begin{gathered}
  {{\dot E}_{{\text{osc}}}} =  - \gamma O\langle \dot \hat x\rangle  =  - O{{\dot J}_O} \hfill \\
  {E_{{\text{int}}}} =  - (\Delta  - 1)O{J_O} =  - \gamma (\Delta  - 1)O\langle \hat x\rangle  \hfill \\
  {M_{{\text{ADM}}}} = \frac{{{\Phi _r}}}{{16\pi G}}M(u) - (\Delta  - 1)O(u){J_O}(u) \hfill \\ 
\end{gathered} \]

So our numerical calculation process in the next part basically follows the following steps:

(1)Use numerical calculation methods to solve the system of equations (30) according to the previous theory.

(2)Without loss of generality, use $\left| {{\psi _0}} \right\rangle ,\left| {{\psi _1}} \right\rangle  \cdots $ Represents the ground state of the quantum state, the first excited state, the second excited state, etc. Use some generalized parameters, $A,{\xi _0}$ (They are just a parameter and should not be understood in a narrow sense as amplitude or phase, etc.) expressing the state in this general form: ${a}(u) = \sum {{A_0}\cos (\omega u + {\xi _0})} $. Use the following formula to find the expected value of each kinematic or dynamic parameter:

$$\langle x(u)\rangle=\langle\psi_{L}(u)|\hat{x}|\psi_{L}(u)\rangle $$

Here $x$ can be $a, b, p_a, p_b$, etc., $u$ represents the variable used.

(3)After studying in this way, the quantum trajectory is mapped to the trajectory on the Bloch sphere. But this way of encoding information is not equidistant. But note that in some specific quantum states (labeled as $\left| 0 \right\rangle ,\left| 1 \right\rangle $), the quantum system and the black hole cannot be coupled. At this time, the quantum system cannot transfer energy to the black hole.

(4)Study the information encoding of the initial state into the black hole. First, in the later stage, the mass of the black hole can be fitted with an exponential function model.

\begin{equation}
\label{deqn_ex1a}
M\left( u \right) = {M_f} - {e^{ - \tau u}}(B\sin \left( {\Omega u + \xi } \right) + C)
\end{equation}

But it is a pity that although this function may be able to obtain a result using the parameter fitting method, the resulting nonlinear response equation of the black hole mass is unsolvable, so a numerical translation is performed in the later stage to eliminate the exponential decay term and the discrete Fourier Transformation processing.

In this way, we have completed the last step, which is how to encode the initial state of the quantum state into the oscillation decay of the black hole (recall that we have explained the encoding mechanism of quantum state evolution in the previous article): Specifically, this initial state is encoded into three initial decay parameters, which are further encoded into Bloch spheres. It is worth noting that the amplitude of the black hole oscillation decay does not vary linearly with $\theta$, but is proportional to the square of $\sin(2\theta)$, while the phase $\xi$ Then there is a linear relationship with negative correlation with $\phi$.

\begin{figure*}[!t]
\centering
\includegraphics[width=6in]{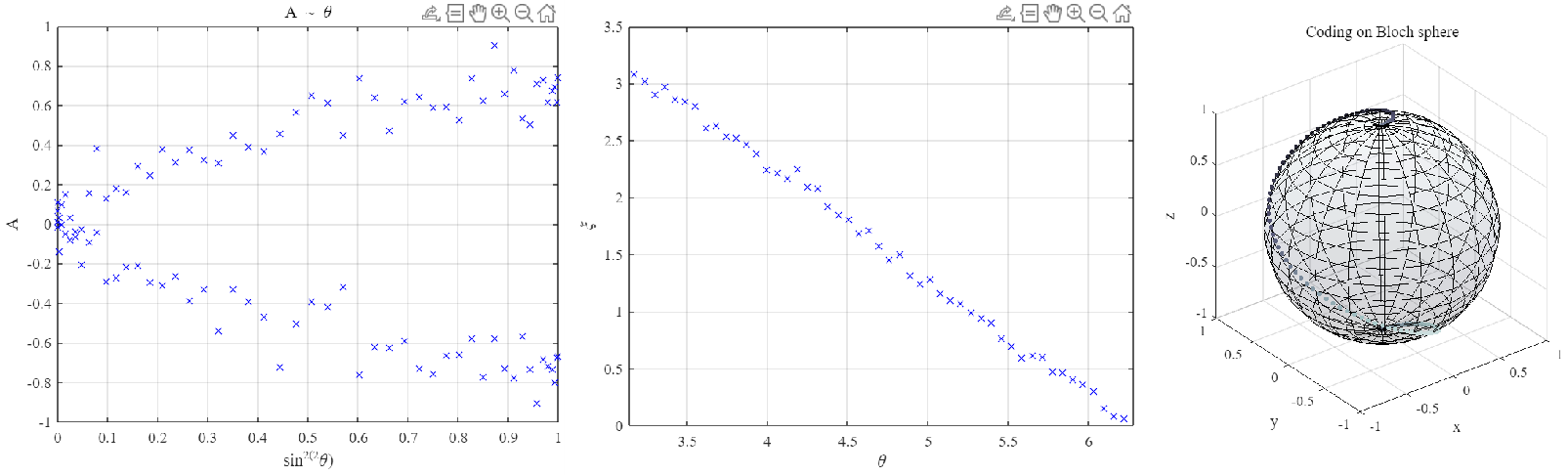}
\caption{Initial state encoding method. Perturbations have been added to this result, two-phase diagrams of three vibration attenuation parameters have been drawn, and the mapped orbit of the quantum trajectory on the Bloch sphere has been projected.}
\label{fig_4}
\end{figure*}

The numerical analyses clearly demonstrate that the parameters \( A \) and \( \xi \) are analogues to the angles \( \theta \) and \( \phi \) on the Bloch sphere, respectively. Consequently, the quantum state's initial configuration is intricately and non-isometrically transcribed into the ringdown phase of the black hole. It is observed that the ringdown amplitude of the black hole does not exhibit a linear correlation; rather, it adheres to a quadratic dependence. Concurrently, the phase \( \xi \) maintains a linear correlation with \( \phi \), akin to \( \chi \), albeit with an inverse gradient. Furthermore, the ringdown amplitude is directly proportional to the energy transference from the oscillator to the black hole's mass.

The time scale $\tau$ of the black hole oscillation decay shows a narrow distribution, and this time scale also encodes the Bloch angle $\phi$. The variation range of $\tau$ $\Delta \tau$ is relative to The ratio of its average value is less than 0.1. This shows that the change amplitude of $\tau$ is very small.

\begin{figure}[!t]
\centering
\includegraphics[width=6in]{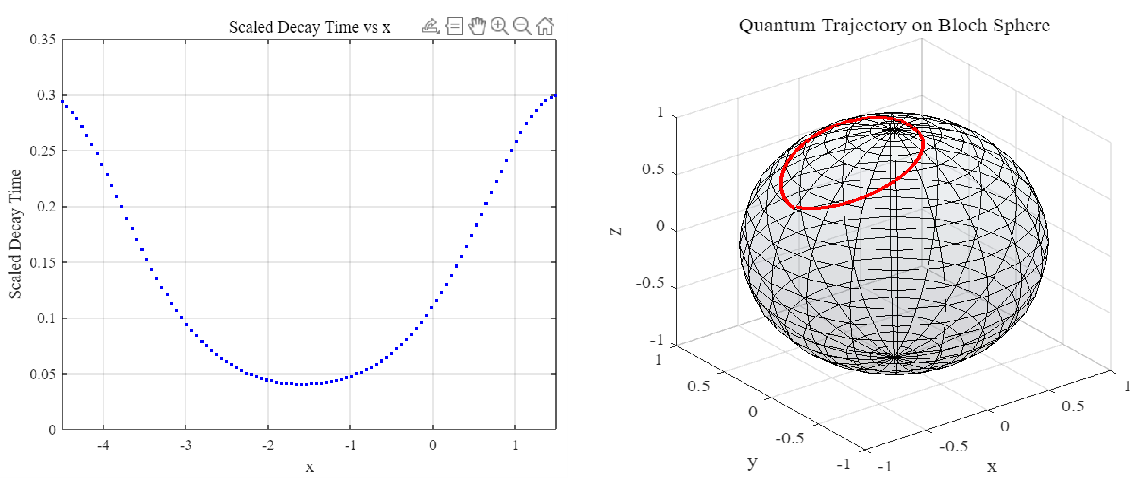}
\caption{The decay exponent also encodes the angle $\phi$ on the Bloch sphere.}
\label{fig_1}
\end{figure}

This coding method is actually applicable when the initial state is a mixed state. for example. Assuming that the oscillation of the quantum state is in the superposition of two ground states, the density evolution matrix of the system satisfies the following expression:

\[\begin{gathered}
  \rho (u = 0) = {\rho _{00}}\left| 0 \right\rangle \left\langle 0 \right| + {\rho _{01}}\left| 0 \right\rangle \left\langle 1 \right| 
   + \rho _{01}^*\left| 1 \right\rangle \left\langle 0 \right| + \left( {1 - {\rho _{00}}} \right)\left| 1 \right\rangle \left\langle 1 \right| \hfill \\ 
\end{gathered} \]

In this scenario, \( \rho_{00} \) is a real entity, whereas \( \rho_{01}^* \) represents the complex conjugate of \( \rho_{01} \). Employing analogous reasoning as applied to the pure state case, it is deduced that the ensuing state constitutes a viable solution to the Liouville equation, which governs the temporal evolution of the density matrix.

\[\begin{gathered}
  \rho (u) = {\rho _{00}}\left| {{\phi _0}(u)} \right\rangle \langle {\phi _0}(u)| + {\rho _{01}}\left| {{\phi _0}(u)} \right\rangle \langle {\phi _1}(u)| \hfill \\
   + \rho _{01}^*\left| {{\phi _1}(u)} \right\rangle \langle {\phi _0}(u)| + \left( {1 - {\rho _{00}}} \right)\left| {{\phi _1}(u)} \right\rangle \langle {\phi _1}(u)| \hfill \\ 
\end{gathered} \]

The transfer of energy from the oscillator to the black hole is:

$$\Delta E=\hbar\omega\rho_{01}\rho_{01}^{*}$$

\begin{figure*}[!t]
\centering
\includegraphics[width=6in]{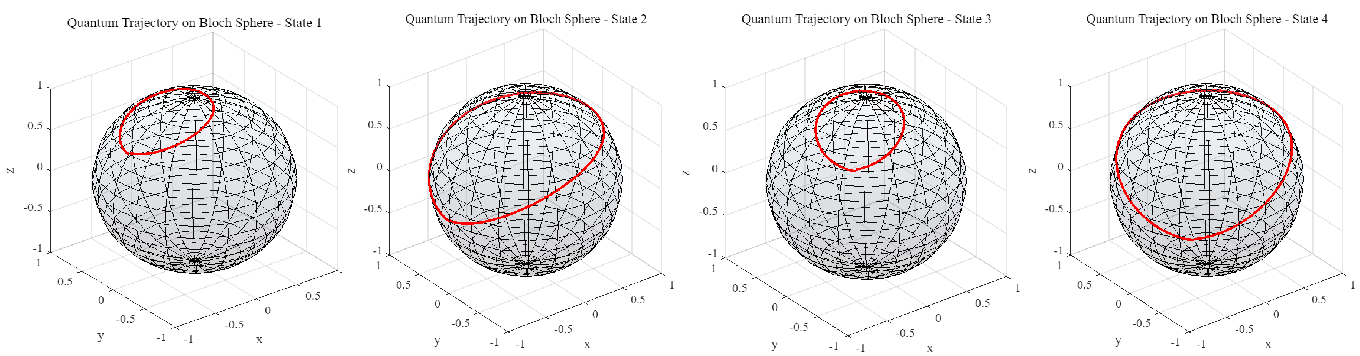}
\caption{Schematic diagram of Bloch sphere encoding method and quantum trajectory on composite initial state}
\label{fig_4}
\end{figure*}

Analogous to the scenario of pure states, the initial state can be inferred from the attenuation of the black hole's oscillatory behavior. The decay parameters \( \tau \) and \( \xi \) encapsulate the Bloch sphere dynamics. Consequently, for any mixed initial state of an oscillator confined within the qubit subspace, its reconstruction is feasible through the analysis of the black hole mass's late-stage oscillatory decay. This approach unveils a crucial element in resolving the information paradox: the state's purity. This attribute exhibits minimal correlation with the oscillation decay's amplitude, phase, or decay rate. In the more general context where the initial state is a superposition encompassing more than two Fock states, its temporal evolution can be efficiently resolved via the invariant method delineated herein. This elucidates that the oscillator's disentanglement from the black hole is a ubiquitous occurrence. Given that the oscillator's evolution adheres to unitary dynamics (more precisely, a nonlinear unitary mean field channel) under the semi-classical large-N approximation, the resultant state preserves the entirety of the initial state's information. Of course, this oversimplification of black holes and quantum information is insufficient to encode all the initial information completely, thus requiring us to develop more refined and analytical models in future research (as we did in the previous part, rather than simply using a model of a quantum oscillator). This can be achieved with quantum computers and quantum optimal control! (For details, please see the discussion in the discussion section).

At the end of this section, we calculate the information entropy of this model. In JT gravity, the entanglement entropy of the state of the dual theory can be holographically computed using the Ryu-Takayanagi formula:\cite{ref38,ref39}

$$S_{\mathrm{ent}}=\frac{\Phi(r_h)}{4G\hbar}$$

The coupling form of two-dimensional JT gravity and material field is:\cite{ref40,ref41,ref42}

\begin{equation}
\label{deqn_ex1a}
\begin{gathered}
  S = \frac{1}{{16\pi G}}\int {{d^2}} x\sqrt { - g} \left( {R + \frac{2}{{{l^2}}}} \right)\Phi  \hfill \\
   + {S_{{\text{matter}}}}[g,\chi ] + \frac{1}{{8\pi G}}\int d u\sqrt { - h} {\phi _b}K + {S_{ct}}[\chi ] \hfill \\
  {S_{{\text{matter}}}} = \int {{{\text{d}}^2}} x\left( { - \frac{1}{2}{g^{\mu \nu }}{\partial _\mu }\chi {\partial _\nu }\chi  - \frac{1}{2}{m^2}{\chi ^2}} \right) \hfill \\ 
\end{gathered} 
\end{equation}

Varying the gravitational action we obtain the following equations of motion:

\begin{equation}
\label{deqn_ex1a}
\begin{aligned}
&R+\frac{2}{l^{2}}=0, \\
&g_{\mu\nu}\left(\nabla^{2}-\frac{1}{l^{2}}\right)\Phi-\nabla_{\mu}\nabla_{\nu}\Phi  =16\pi G\left(\partial_{\mu}\chi\partial_{\nu}\chi-\frac{1}{2}g_{\mu\nu}g^{\alpha\beta}\partial_{\alpha}\chi\partial_{\beta}\chi-\frac{1}{2}g_{\mu\nu}m^{2}\chi^{2}\right) \\
&(\nabla^{2}-m^{2})\chi=0
\end{aligned}
\end{equation}

Expression transformation of metric:

$$\mathrm{d}s^2=-2\frac{l^2}{r^2}\mathrm{d}u\mathrm{d}r-\left(\frac{l^2}{r^2}-M(u)l^2\right)\mathrm{d}u^2$$

Make:

\[\left\{ \begin{gathered}
  T = T(u),\quad R = \frac{{T'(u)r}}{{1 - \frac{{T''(u)}}{{T'(u)}}r}} \hfill \\
   - 2Sch(T(u),u) + T'{(u)^2} = M(u) \hfill \\ 
\end{gathered}  \right.\]

Transforming the above two homomorphisms into dynamic horizons, we can get:\cite{ref43}

\begin{equation}
\label{deqn_ex1a}
r_h(u)=\frac l{T^{\prime}(u)\left(1+\frac{T^{\prime\prime}(u)}{T^{\prime}(u)^2}\right)}
\end{equation}

When considering the evolution of the black hole horizon, a smooth transition occurs between the thermal horizon from the initial to the final state, and by applying the solution of the diluton equation:

\[\begin{gathered}
  \Phi  = \frac{{{\Phi _r}}}{r}
   + 16\pi G\left( {\frac{{{\chi ^{(0)}}{{(u)}^2}}}{4}{r^{ - 1/2}} + \int_0^r d r''\frac{1}{{{{r''}^2}}}} \right. 
  \left. {\left( {\frac{1}{8}{\chi ^{(0)}}{{(u)}^2}{r^{\prime \prime \frac{1}{2}}} - \int_0^{r''} d r'{{r'}^2}{{\left( {{\partial _{r'}}\chi (r',u)} \right)}^2}} \right)} \right) \hfill \\ 
\end{gathered} \]

The entropy of the black hole exhibits a monotonically increasing trend over time, consistent with the second law of thermodynamics, which posits that the total entropy always tends to ascend. In the context of the semiclassical mean field approximation, wherein the oscillator's evolution remains unitary and the relevant volume scalars assume classical characteristics, the consideration of alternative forms of time-dependent entropy becomes unnecessary. The observed monotonic growth of entropy highlights a pivotal phenomenon: within the semiclassical throat geometry of a black hole, information undergoes an irreversible loss. Although the process of black hole oscillation decay temporarily encodes information regarding the initial state, this information eventually finds encoding in Hawking radiation. As elaborated in the subsequent section, the energy loss attributable to Hawking radiation is directly proportional to the black hole mass \(M(u)\). This implies that meticulous measurements of the energy flux of Hawking radiation can unveil the characteristics of the qubit's initial state.

\begin{figure*}[!t]
\centering
\includegraphics[width=5in]{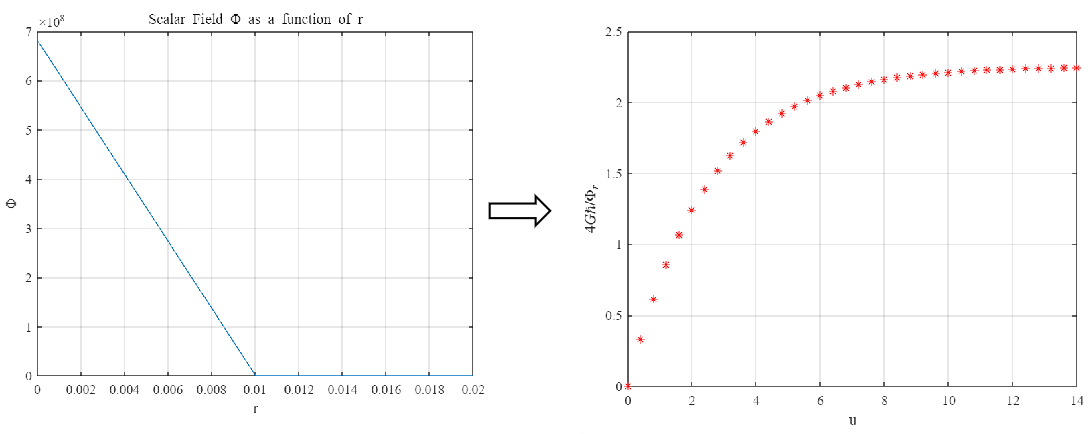}
\caption{The evolution process of information entropy}
\label{fig_4}
\end{figure*}

A brief summary of the work performed in this subsection: We explore the information encoding and transfer mechanisms and information loss mechanisms in black holes. For more precise research, we still use the simplified model of the quantum resonator for the quantum system. For the black hole model, we use the precise theoretical model established in section II. For the transmission mechanism of quantum information in the black hole, we use various dynamic parameters is represented by the evolution equation, and the encoding form of the initial state (whether it is a pure state or a superposition state) is characterized by a Bloch sphere. The loss of pure states is a quantum information paradox.

\subsection{The decoding mechanism of information in black holes}

The previous part introduced the mechanism of encoding material information into black holes and information propagation. To briefly repeat, a part of the energy is coupled with the black hole (this part of the information propagation process is irreversible), and the time-varying state of the remaining quantum matter The initial state of matter is encoded in a non-equidistant manner, while the oscillation process in the throat encodes another non-equidistant, temporary copy of the incoming matter. This copy will enter Hawking radiation, the process of information leaving the black hole. Then "evaporate" in Hawking radiation.

Use Poincaré coordinates to describe the metric:

\[\begin{gathered}
  d{s^2} =  - \frac{{4d{x^ + }d{x^ - }}}{{{{({x^ + } - {x^ - })}^2}}} = \frac{{4dxd\bar x}}{{{{(x + \bar x)}^2}}} \hfill \\
  {\text{with}},\quad {x^ + } = t + z = \bar x,{x^ - } = t - z =  - x \hfill \\ 
\end{gathered} \]

By using diffeomorphisms, the initial state of matter is mapped to a specific vacuum state. Certain expected values of physical quantities remain constant in the coordinate system considered. There is the following equation:

$$
\left(\frac{dw}{dx}\right)^2\langle T_{ww}\rangle=\langle T_{xx}\rangle+\frac{c}{24\pi}Sch(w(x),x)
$$

The difepmorphism is the following: \cite{ref44}

\begin{equation}
\label{deqn_ex1a}
w(x)=\begin{cases}\frac{w_0^2\dot t(0)}{x+t(0)w_0},&\quad x>0\\w_0+t^{-1}(-x),&\quad x<0\end{cases}
\end{equation}

It is necessary to introduce the structural form of the JT gravity model in the AdS2 space under the single throat model. This is a microscopic model of a black hole.

\begin{equation}
\label{deqn_ex1a}
\begin{aligned}
&\mathcal{Q}_{i}^{+} =\frac{\ddot{t}_{i}(u)}{\dot{t}_{i}(u)^{2}}-\frac{\ddot{t}_{i}(u)^{2}}{\dot{t}_{i}(u)^{3}},  \\
&\mathcal{Q}_{i}^{0} =t_{i}(u)\left(\frac{\ddot{t}_{i}(u)}{\dot{t}_{i}(u)^{2}}-\frac{\ddot{t}_{i}^{2}}{\dot{t}_{i}(u)^{3}}\right)-\frac{\ddot{t}_{i}(u)}{\dot{t}_{i}(u)},  \\
&\mathcal{Q}_i^- =t_{i}(u)^{2}\left(\frac{\ddot{t}_{i}(u)}{\dot{t}_{i}(u)^{2}}-\frac{\ddot{t}_{i}(u)^{2}}{\dot{t}_{i}(u)^{3}}\right)-2t_{i}(u)\left(\frac{\ddot{t}_{i}(u)}{\dot{t}_{i}(u)}-\frac{\dot{t}_{i}(u)}{t_{i}(u)}\right) 
\end{aligned}
\end{equation}

where $t$ is the renormalization parameter. The equivalent construction format of the mass evolution equation:

$$M_i=-\mathcal{Q}_i\cdot\mathcal{Q}_i=-2Sch(t_i(u),u)$$

Quantum dots obey the discrete Klein-Gordon equation:

\begin{equation}
\label{deqn_ex1a}
\ddot{\mathbf{Q}}_i-\frac{1}{\sigma^2}(\mathbf{Q}_{i-1}+\mathbf{Q}_{i+1}-2\mathbf{Q}_i)=0
\end{equation}

Therefore, the dynamic equations of the microscopic model are: \cite{ref45,ref46}

\begin{equation}
\label{deqn_ex1a}
\left\{ \begin{gathered}
  {{\dot M}_i} =  - \lambda \left( {{{\mathbf{Q}}_{i - 1}} + {{\mathbf{Q}}_{i + 1}} - 2{{\mathbf{Q}}_i}} \right) \cdot {\mathbf{Q}}_i^\prime  \hfill \\
  {\mathop {\mathbf{Q}}\limits^{..} _i} = \frac{1}{{{\sigma ^2}}}\left( {{{\mathbf{Q}}_{i - 1}} + {{\mathbf{Q}}_{i + 1}} - 2{{\mathbf{Q}}_i}} \right) + \frac{1}{{{\lambda ^2}}}\left( {{{\mathbf{Q}}_{i - 1}} + {{\mathbf{Q}}_{i + 1}} - 2{{\mathbf{Q}}_i}} \right) \hfill \\
  E = {E_\mathcal{Q}} + {E_{\mathbf{Q}}} \hfill \\
  {E_\mathcal{Q}} = \sum\limits_i {{M_i}}  = M \hfill \\
  {E_{\mathbf{Q}}} = \frac{{{\lambda ^3}}}{2}\sum\limits_i {{{\mathop {\mathbf{Q}}\limits^. }_i}}  \cdot {\mathop {\mathbf{Q}}\limits^. _i} + \frac{{{\lambda ^3}}}{{2{\sigma ^2}}}\sum\limits_i {({{\mathbf{Q}}_{i + 1}} - {{\mathbf{Q}}_i})}  \cdot ({{\mathbf{Q}}_{i + 1}} - {{\mathbf{Q}}_i}) \hfill \\ 
\end{gathered}  \right.
\end{equation}

According to the theory at the beginning of this section, the two terms of this system of equations are modified to:

$$\begin{aligned}
&\dot{M}_{i} =-\lambda({\mathcal Q}_{i-1}+{\mathcal Q}_{i+1}-2{\mathcal Q}_{i})\cdot\dot{\mathbf Q}_{i}-\frac{c}{24\pi}M_{i},  \\
&\ddot{\mathbf{Q}}_{i} =\frac{1}{\sigma^{2}}(\mathbf{Q}_{i-1}+\mathbf{Q}_{i+1}-2\mathbf{Q}_{i})+\frac{1}{\lambda^{2}}(\mathbf{Q}_{i-1}+\mathbf{Q}_{i+1}-2\mathbf{Q}_{i}) 
\end{aligned}$$

To extract the decoupling time scale, we consider:

$$D_i\equiv\dot{M}_i+\frac{c}{24\pi}M_i=-\lambda(\mathcal{Q}_{i-1}+\mathcal{Q}_{i+1}-2\mathcal{Q}_i)\cdot\dot{\mathbf{Q}}_i$$

Although this model is based on the information decoding mechanism proposed by string theory (single throat model), it does not seem to be related to the LQG theory used to describe the evolution of quantum information that we have discussed previously, but we might as well take a look at the information decoding mechanism from string theory How to get this conclusion.

The overall decoupling process is described in one paragraph. As the information leaves, the mass of all MDPs decays to 0, the energy decays to 0, and the Noether electrons on all quantum dots also decay to 0. The paradox of the resulting information loss is that because this encoding method is uniform, all information is lost in the process. But we can still realize the evolution process of energy through ring quantum theory. And we speculate that this loss of information is reflected by the non-conservation of energy before and after.

\section{Results}

\subsection{Advantages of Ring Quantum Theory Model in Simulating Black Hole Quantum Dynamics}

At the beginning of the results section, we solve the kinematic equations using different effective Hamiltonians. Mainly by comparing the differences in solutions to the equations of motion under different initial conditions and initial parameters, and we focus on some singular points. , or the "white hole" concept introduced by ring quantum theory, etc. Substitute equations (21) (22) (24) - (26) into the following equations:

$$
\begin{gathered}
  \dot a(T) = \{ a(T),H_{{\text{eff}}}^{(i)}\} , \hfill \\
  \dot b(T) = \{ b(T),H_{{\text{eff}}}^{(i)}\} , \hfill \\
  {{\dot p}_a}(T) = \{ {p_a}(T),H_{{\text{eff}}}^{(i)}\} , \hfill \\
  {{\dot p}_b}(T) = \{ {p_b}(T),H_{{\text{eff}}}^{(i)}\}  \hfill \\ 
\end{gathered} 
$$

As mentioned before, the analytical solution of the equations of motion requires the selection of a suitable attenuation function. We choose the symbolic function: $N=\mathrm{sgn}(p_{b})\sqrt{|p_{a}|}$. In order to compare with existing research results, we use the same initial conditions:

$$a^0=const,\quad b^0=0,\quad p_a^0=4M_{\mathrm{bh}}^2,\quad p_b^0=0$$

The most important parameter of the initial conditions is $a^0$, which reproduces the Schwarzschild solution on the horizon, allowing us to interpret the resulting dynamics as the (effective) evolution of the interior of the Schwarzschild black hole. This determines different evolutionary forms. Another analysis is about the analysis of quantization parameters. The quantization scheme used in equations (24)~(26) here is $\mu_0$-scheme. Such as $\mu_{b}=\sqrt{\Delta/p_{a}^{0}}$ \cite{ref47} Under this method, the choice of quantization parameters is independent of the initial conditions. But in our other analytical expansion expression, the $bar \mu $ method is used. The latter method is considered to be more "physical". The first theoretical formula is:

$$\mu_1=\mu_a=\mu_a^o,\quad\mu_2=\mu_3=\mu_b=\mu_b^o,$$

Therefore, this theory is suitable for us to analyze the phase diagram (or quantum trajectory diagram) of each kinematic parameter and change the initial conditions, and the results obtained are as shown in the figure 7 to figure 10

\begin{figure*}[!t]
\centering
\includegraphics[width=6in]{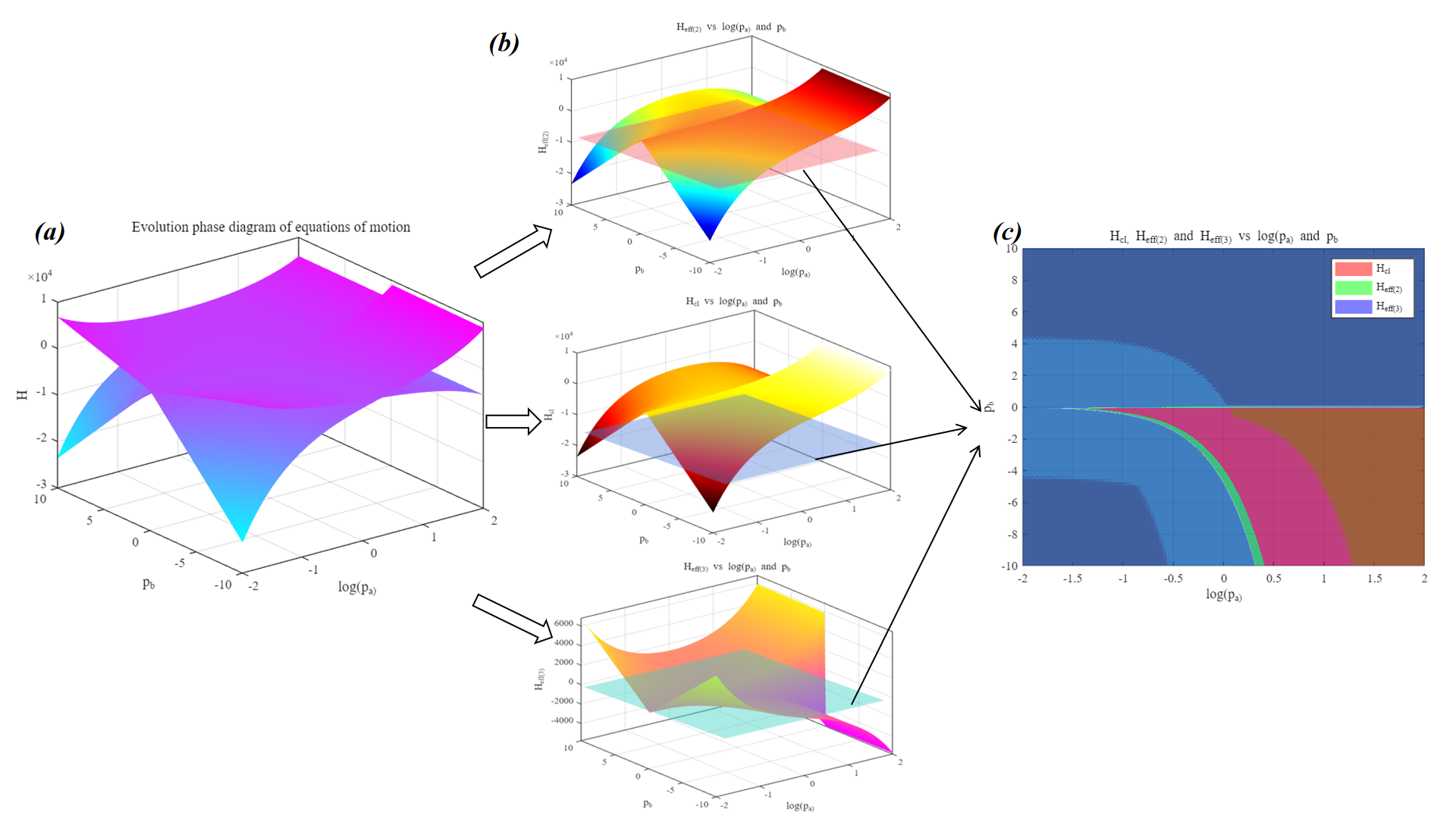}
\caption{Condition: Evolution equation obtained by $G=c=1,\mu_a=\mu_b=1/10,\beta=0.2375,a=b=1,N=1$. (a) Image of three trajectory phase diagrams superimposed (b) Three trajectory phase diagrams Figure separated image (c) projection image.}
\label{fig_4}
\end{figure*}

\begin{figure*}[!t]
\centering
\includegraphics[width=6in]{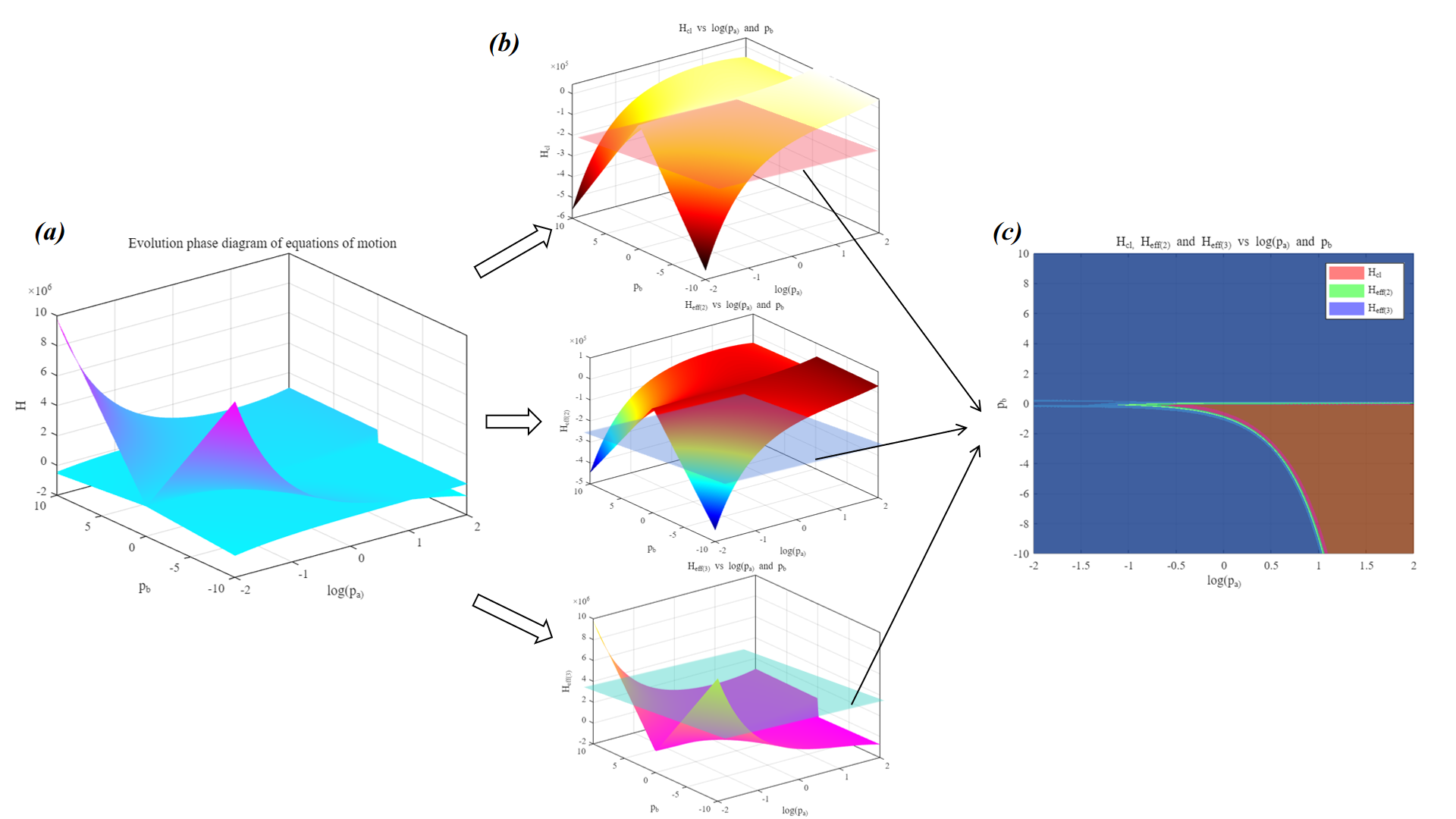}
\caption{Condition: Evolution equation obtained by $G=c=1,\mu_a=\mu_b=1/10,\beta=0.2375,a=1,b=5.N=1$. (a) Image of three trajectory phase diagrams superimposed (b) Three trajectory phase diagrams Figure separated image (c) projection image.}
\label{fig_4}
\end{figure*}

\begin{figure*}[!t]
\centering
\includegraphics[width=6in]{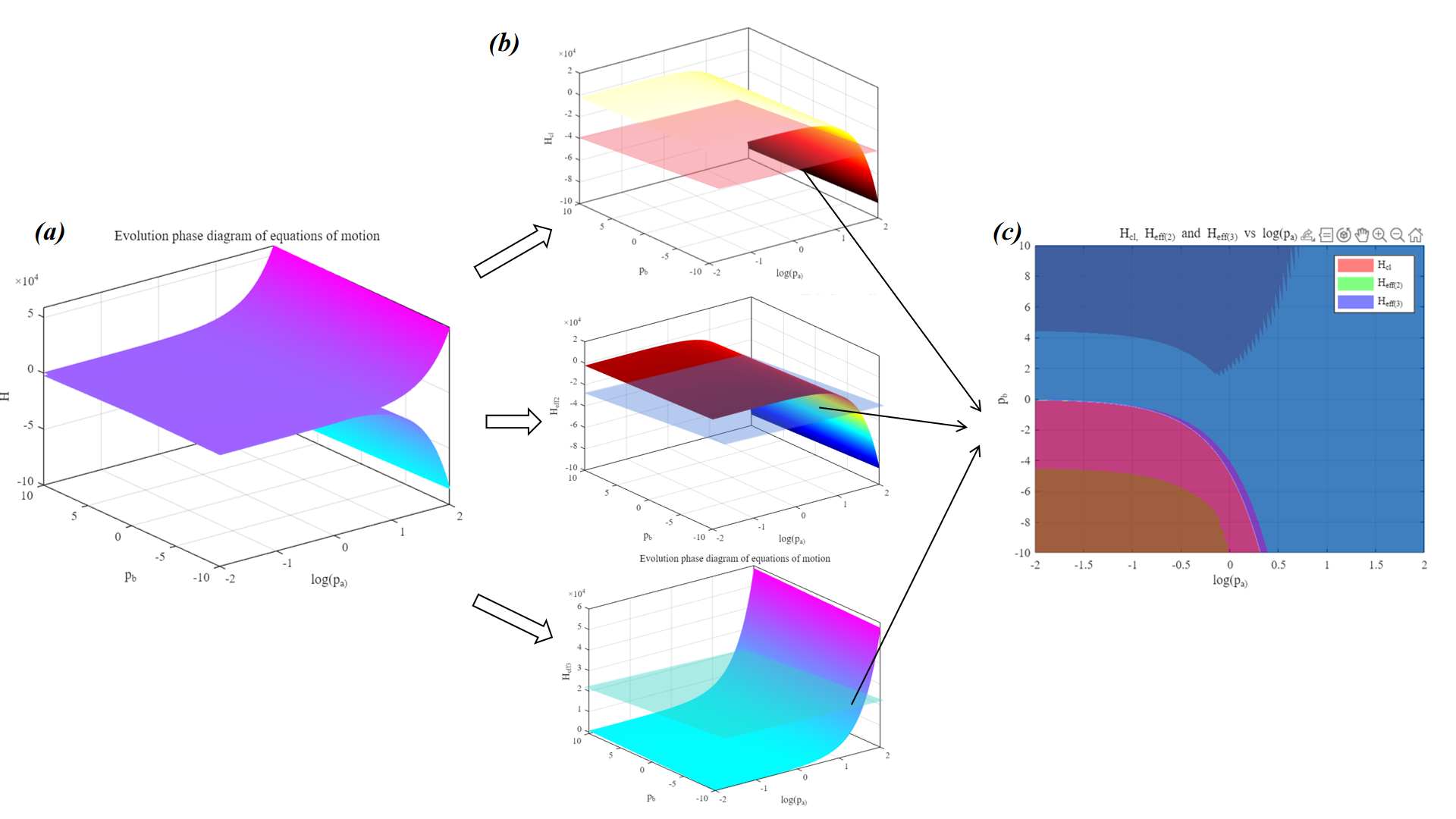}
\caption{Condition: Evolution equation obtained by $G=c=1,\mu_a=\mu_b=1/10,\beta=0.2375,a=1,b=1,N=\mathrm{sgn}(p_{b})\sqrt{|p_{a}|}$. (a) Image of three trajectory phase diagrams superimposed (b) Three trajectory phase diagrams Figure separated image (c) projection image.}
\label{fig_4}
\end{figure*}

\begin{figure*}[!t]
\centering
\includegraphics[width=6in]{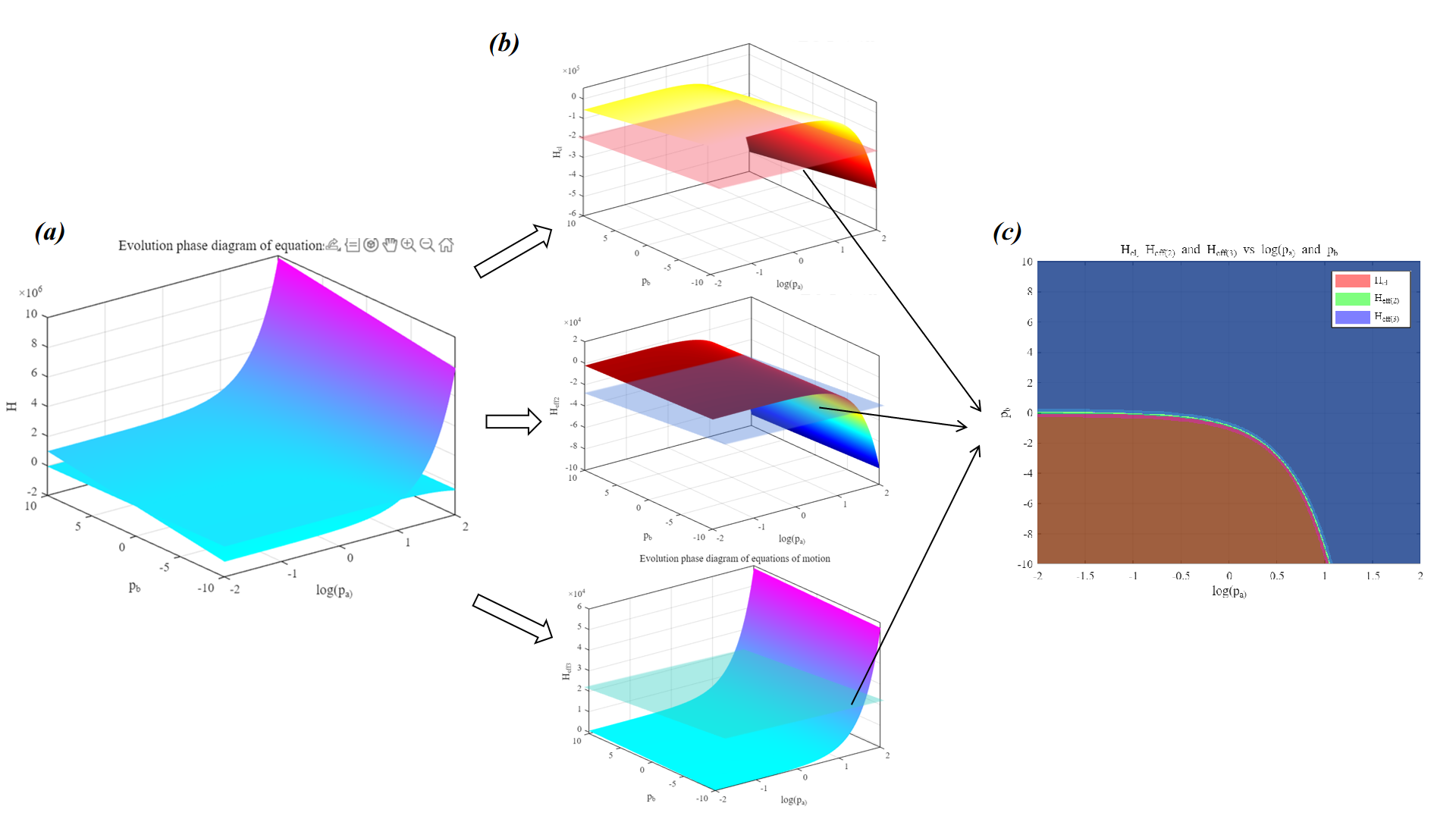}
\caption{Condition: Evolution equation obtained by $G=c=1,\mu_a=\mu_b=1/10,\beta=0.2375,a=1,b=5,N=\mathrm{sgn}(p_{b})\sqrt{|p_{a}|}$. (a) Image of three trajectory phase diagrams superimposed (b) Three trajectory phase diagrams Figure separated image (c) projection image.}
\label{fig_4}
\end{figure*}

The following calculation results are the calculation results when the initial condition $a$ is set to 1 and 5 respectively, and the attenuation function $N$ is set to a constant 1 and a sign function respectively. During the calculation process, we carried out first-order, second-order and third-order expansions of the precise analytical formula of the Hamiltonian, and drew the quantum trajectory diagram. Note that only the effective values are shown in the figure, that is, $p_a, p_b$ is not 0 Effective solution. (a) and (b) of each diagram are three-dimensional phase diagrams, and (c) is the projected two-dimensional phase diagram. It can be seen from the figure: In the first-order and second-order expansions, we have not found the position that can be used to represent the singular point (or jump point, white hole point) in the evolution process, and the first- and second-order expansions The results are very similar. Starting from the third level of expansion, this difference can be clearly seen. The specific performance is as follows (mainly see picture (c)): starting from the third-order expansion, there is a vanishing point of $p_b$ (before this, it was a bottoming out situation, which means there is no predicted singularity), It shows that there is a killing hand horizon in the effective solution, and it eventually converges to a solution that is completely different from the first-order and second-order expansions.

By evaluating the geodesic expansions, denoted as \(\Theta_{\pm}\), associated with the two future-oriented null normals to the surfaces of constant \(T\) and \(R\), it is noteworthy that these quantities exhibit a consistent sign throughout the evolution. They are directly related to the sign of \(\displaystyle \frac{\dot{p}_a}{N}\), a parameter that remains invariant as \(p_a\) maintains a positive value during the entire evolution process. It is important to emphasize that this relationship remains independent of the choice of coordinate systems. Specifically, within the region \(0 < T < T_b\), \(\displaystyle \frac{\dot{p}_a}{N}\) assumes a negative value, while in the region \(T_b < T < T_{wh}\), it is positive. Moreover, at the transition point \(T_b\), this parameter vanishes. Consequently, the boundary surface \(T = T_b\) can be characterized as a transitional boundary demarcating a shift from a trapped region – representing the interior of the black hole – to an anti-trapped region, akin to the interior of a 'white hole.'

Further solve the evolution equation and recall the following corresponding coordinate mapping:

\[\begin{gathered}
  E_1^1 = {p_a}\sin (\theta ),E_2^2 = {p_b}\sin (\theta ),E_3^3 = {p_b}, \hfill \\
  A_1^1 = 2\beta {K_a},A_2^2 = \beta {K_b},A_3^3 = \beta {K_b}\sin (\theta ) \hfill \\
  A_3^1 = \cos (\theta ),A_3^2 =  - \sin (\theta )\frac{{{p_a}}}{{2{p_b}}},A_2^3 = \frac{{{p_a}}}{{2{p_b}}} \hfill \\ 
\end{gathered} \]

And the following exchange relationship:\cite{ref48}

\[\begin{gathered}
  \left\{ {{K_a}(x),{p_a}\left( {x'} \right)} \right\} = G\delta \left( {x - x'} \right) \hfill \\
  \left\{ {{K_b}(x),{p_b}\left( {x'} \right)} \right\} = G\delta \left( {x - x'} \right) \hfill \\ 
\end{gathered} \]

With the Ashtekar-Barbero connection: $\displaystyle A_1^1 = 2\beta {K_a},A_2^2 = \beta {K_b},A_3^3 = \beta {K_b}\sin \theta $ The LQG theoretical writing:

\[{K_b} \to \frac{{\sqrt {|{p_a}|} }}{\beta }\sin \left[ {\frac{{\beta {p_b}}}{{\sqrt {|{p_a}|} }}} \right],{K_a} \to \frac{{{p_b}}}{{2\beta \sqrt {|{p_a}|} }}\sin \left[ {\frac{{\sqrt {|{p_a}|} }}{{{p_b}}}2\beta {p_a}} \right]{\text{ }}\]

Make the following quantum approximation:

\[\frac{{8\pi {p_a}\delta {p_a}}}{{\kappa \sqrt {|{p_a}|} |{p_b}|}},\quad  - \frac{{8\pi {p_a}\delta {p_b}|{p_b}|{p_a}}}{{\kappa p_b^3\sqrt {|{p_a}|} }}.\]

The next step is to solve the time-dependent and spatial evolution rates of the Hamiltonian at different analytical precisions. In order to avoid the insolvability of directly substituting into the Schrödinger equation, we perform the following operations to first convert it from a partial differential equation into a system of ordinary differential equations.

\[\begin{gathered}
  {\partial _t}{K_a} =  - \frac{{{\partial _a}{p_a}{\partial _b}{p_b}}}{{4\sqrt {{p_a}} p_b^2}} - \frac{{{{({\partial _a}{p_a})}^2}}}{{16p_a^{3/2}{p_b}}} + \frac{{\partial _a^2{p_a}}}{{4\sqrt {{p_a}} {p_b}}} + \frac{{{p_b}}}{{4p_a^{3/2}}} 
   - \frac{{{p_b}\sin \left( {\frac{{\beta {K_b}}}{{\sqrt {{p_a}} }}} \right)\sin \left( {\frac{{2\beta \sqrt {{p_a}} {K_a}}}{{{p_b}}}} \right)}}{{2{\beta ^2}{p_a}}} \hfill \\
   - \frac{{{K_a}\sin \left( {\frac{{\beta {K_b}}}{{\sqrt {{p_a}} }}} \right)\cos \left( {\frac{{2\beta \sqrt {{p_a}} {K_b}}}{{{p_b}}}} \right)}}{\beta }
   + \frac{{{p_b}{K_\varphi }\cos \left( {\frac{{\beta {K_b}}}{{\sqrt {{p_a}} }}} \right)\sin \left( {\frac{{2\beta \sqrt {{p_a}} {K_b}}}{{{p_b}}}} \right)}}{{2\beta {p_a}}} - \frac{{{p_b}{{\sin }^2}\left( {\frac{{\beta {K_b}}}{{\sqrt {{p_a}} }}} \right)}}{{4{\beta ^2}{p_a}}} \hfill \\
   + \frac{{{p_b}{K_b}\sin \left( {\frac{{\beta {K_b}}}{{\sqrt {{p_a}} }}} \right)\cos \left( {\frac{{\beta {K_b}}}{{\sqrt {{p_a}} }}} \right)}}{{2\beta {p_a}}} \hfill \\ 
\end{gathered} \]

\[\begin{gathered}
  {\partial _t}{K_b} = \frac{{{{({\partial _x}{p_a})}^2}}}{{8\sqrt {{p_a}} p_b^2}} - \frac{{\sqrt {{p_a}} \sin \left( {\frac{{\beta {K_b}}}{{\sqrt {{p_a}} }}} \right)\sin \left( {\frac{{2\beta \sqrt {{p_a}} {K_\chi }}}{{{p_b}}}} \right)}}{{{\beta ^2}}}
   + \frac{{2{p_a}{K_a}\sin \left( {\frac{{\beta {K_b}}}{{\sqrt {{p_a}} }}} \right)\cos \left( {\frac{{2\beta \sqrt {{p_a}} {K_a}}}{{{p_b}}}} \right)}}{{\beta {p_b}}} \hfill \\
   - \frac{{\sqrt {{p_a}} {{\sin }^2}\left( {\frac{{\beta {K_b}}}{{\sqrt {{p_a}} }}} \right)}}{{2{\beta ^2}}} - \frac{1}{{2\sqrt {{p_a}} }}, \hfill \\
  {\partial _t}{p_a} = \frac{{2{p_a}\sin \left( {\frac{{\beta {K_b}}}{{\sqrt {{p_a}} }}} \right)\cos \left( {\frac{{2\beta \sqrt {{p_a}} {K_\chi }}}{{{p_b}}}} \right)}}{\beta }, \hfill \\
  {\partial _t}{p_b} = \frac{{{p_b}\cos \left( {\frac{{\beta {K_b}}}{{\sqrt {{p_a}} }}} \right)\sin \left( {\frac{{2\beta \sqrt {{p_a}} {K_\chi }}}{{{p_b}}}} \right)}}{\beta } 
   + \frac{{{p_b}\sin \left( {\frac{{\beta {K_b}}}{{\sqrt {{p_a}} }}} \right)\cos \left( {\frac{{\beta {K_b}}}{{\sqrt {{p_a}} }}} \right)}}{\beta } \hfill \\ 
\end{gathered} \]

The form of the dual equation is as follows\cite{ref49,ref50,ref51,ref52},Satisfying the following dual relationship:

$$\tilde{x}=-x,\quad\tilde{t}=-t$$

with the fields transforming as follows:

\[\tilde K(\tilde t,\tilde x) = \tilde K( - t, - x) = {K_x}(t,x)\]

\[\begin{gathered}
  {\partial _{\tilde t}}{K_{\tilde a}} =  - \frac{{{\partial _{\tilde a}}{p_{\tilde a}}{\partial _{\tilde b}}{p_{\tilde b}}}}{{4\sqrt { - {p_{\tilde a}}} {p_{{{\tilde b}^2}}}}} + \frac{{{{({\partial _{\tilde a}}{p_{\tilde a}})}^2}}}{{16{{( - {p_{\tilde a}})}^{3/2}}{p_{\tilde b}}}} + \frac{{\partial _{\tilde a}^2{p_{\tilde a}}}}{{4\sqrt { - {p_{\tilde a}}} {p_{\tilde b}}}} 
   - \frac{{{p_{\tilde b}}}}{{4{{( - {p_{\tilde a}})}^{3/2}}}} - \frac{{{p_{\tilde b}}\sin \left( {\frac{{\beta {K_{\tilde b}}}}{{\sqrt { - {p_{\tilde a}}} }}} \right)\sin \left( {\frac{{2\beta \sqrt { - {p_{\tilde a}}} {K_{\tilde a}}}}{{{p_{\tilde b}}}}} \right)}}{{2{\beta ^2}\sqrt { - {p_{\tilde a}}} }} \hfill \\
   - \frac{{{K_{\tilde a}}\sin \left( {\frac{{\beta {K_{\tilde b}}}}{{\sqrt { - {p_{\tilde a}}} }}} \right)\cos \left( {\frac{{2\beta \sqrt { - {p_{\tilde a}}} {K_{\tilde a}}}}{{{p_{\tilde b}}}}} \right)}}{\beta }
   - \frac{{{p_{\tilde b}}{K_{\tilde b}}\cos \left( {\frac{{\beta {K_{\tilde b}}}}{{\sqrt { - {p_{\tilde a}}} }}} \right)\sin \left( {\frac{{2\beta \sqrt { - {p_{\tilde a}}} {K_{\tilde a}}}}{{{p_{\tilde b}}}}} \right)}}{{2\beta {p_{\tilde a}}}} \hfill \\
   + \frac{{{p_{\tilde b}}{{\sin }^2}\left( {\frac{{\beta {K_{\tilde b}}}}{{\sqrt { - {p_{\tilde a}}} }}} \right)}}{{4{\beta ^2}\sqrt { - {p_{\tilde a}}} }} + \frac{{{p_{\tilde b}}{K_{\tilde b}}\sin \left( {\frac{{\beta {K_{\tilde b}}}}{{\sqrt { - {p_{\tilde a}}} }}} \right)\cos \left( {\frac{{\beta {K_{\tilde b}}}}{{\sqrt { - {p_{\tilde a}}} }}} \right)}}{{2\beta {p_{\tilde a}}}} \hfill \\ 
\end{gathered} \]

\[\begin{gathered}
  {\partial _t}{K_{\bar b}} = \frac{{{{({\partial _{\tilde a}}{p_{\tilde a}})}^2}}}{{8\sqrt { - {p_{\tilde a}}} {p_{{{\tilde b}^2}}}}} + \frac{{\sqrt { - {p_{\tilde a}}} \sin \left( {\frac{{\beta {K_{\tilde b}}}}{{\sqrt { - {p_{\tilde a}}} }}} \right)\sin \left( {\frac{{2\beta \sqrt { - {p_{\tilde a}}} {K_{\tilde a}}}}{{{p_{\tilde b}}}}} \right)}}{{{\beta ^2}}}
   + \frac{{2{p_{\tilde a}}{K_{\tilde a}}\sin \left( {\frac{{\beta {K_{\tilde b}}}}{{\sqrt { - {p_{\tilde a}}} }}} \right)\cos \left( {\frac{{2\beta \sqrt { - {p_{\tilde a}}} {K_{\tilde a}}}}{{{p_{\tilde b}}}}} \right)}}{{\beta {p_{\tilde b}}}} \hfill \\
   - \frac{{\sqrt { - {p_{\tilde a}}} {{\sin }^2}\left( {\frac{{\beta {K_{\tilde b}}}}{{\sqrt { - {p_{\tilde a}}} }}} \right)}}{{2{\beta ^2}}} - \frac{1}{{2\sqrt { - {p_{\tilde a}}} }} \hfill \\
  {\partial _t}{p_{\tilde a}} = \frac{{2{p_{\bar a}}\sin \left( {\frac{{\beta {K_{\bar b}}}}{{\sqrt { - {p_{\bar a}}} }}} \right)\cos \left( {\frac{{2\beta \sqrt { - {p_{\bar a}}} {K_{\bar a}}}}{{{p_{\bar b}}}}} \right)}}{\beta }, \hfill \\
  {\partial _t}{p_{\tilde b}} = \frac{{{p_{\hat b}}\sin \left( {\frac{{\beta {K_{\tilde b}}}}{{\sqrt { - {p_{\tilde a}}} }}} \right)\cos \left( {\frac{{\beta {K_{\tilde b}}}}{{\sqrt { - {p_{\tilde a}}} }}} \right)}}{\beta } 
   - \frac{{{p_{\tilde b}}\cos \left( {\frac{{\beta {K_{\tilde b}}}}{{\sqrt { - {p_{\tilde a}}} }}} \right)\sin \left( {\frac{{2\beta \sqrt { - {p_{\tilde a}}} {K_{\tilde a}}}}{{{p_{\tilde b}}}}} \right)}}{\beta } \hfill \\ 
\end{gathered} \]

Therefore, the Schrödinger equation can be transformed into the following set of ordinary differential equations:

\[\begin{gathered}
  \frac{{\text{d}}}{{{\text{d}}t}}\left[ {\begin{array}{*{20}{c}}
  {{p_a}} \\ 
  {{p_b}} \\ 
  {{K_a}} \\ 
  {{K_b}} 
\end{array}} \right] = \left[ {\begin{array}{*{20}{c}}
  {{f^a}({p_a},{p_b},{K_a},{K_b})} \\ 
  {{f^b}({p_a},{p_b},{K_a},{K_b})} \\ 
  {{f_1}({p_a},{p_b},{K_a},{K_b})} \\ 
  {{f_2}({p_a},{p_b},{K_a},{K_b})} 
\end{array}} \right] \hfill \\
  \frac{{\text{d}}}{{{\text{d}}\tilde t}}\left[ {\begin{array}{*{20}{c}}
  {{p_{\tilde a}}} \\ 
  {{p_{\tilde b}}} \\ 
  {{{\tilde K}_a}} \\ 
  {{{\tilde K}_b}} 
\end{array}} \right] = \left[ {\begin{array}{*{20}{c}}
  {{f^{\tilde a}}({p_{\tilde a}},{p_{\tilde b}},{{\tilde K}_a},{{\tilde K}_b})} \\ 
  {{f^{\tilde b}}({p_{\tilde a}},{p_{\tilde b}},{{\tilde K}_a},{{\tilde K}_b})} \\ 
  {{{\tilde f}_1}({p_{\tilde a}},{p_{\tilde b}},{{\tilde K}_a},{{\tilde K}_b})} \\ 
  {{{\tilde f}_2}({p_{\tilde a}},{p_{\tilde b}},{{\tilde K}_a},{{\tilde K}_b})} 
\end{array}} \right] \hfill \\ 
\end{gathered} \]

The schematic diagram of the evolution of this system of equations will be placed in Figure 11. This schematic diagram of evolution is mainly used to solve the system of equations of energy transfer and transformation in the next section. The results here will not be used to explain the comparative differences after the expansion of each level of refined solutions.

\begin{figure*}[!t]
\centering
\includegraphics[width=6in]{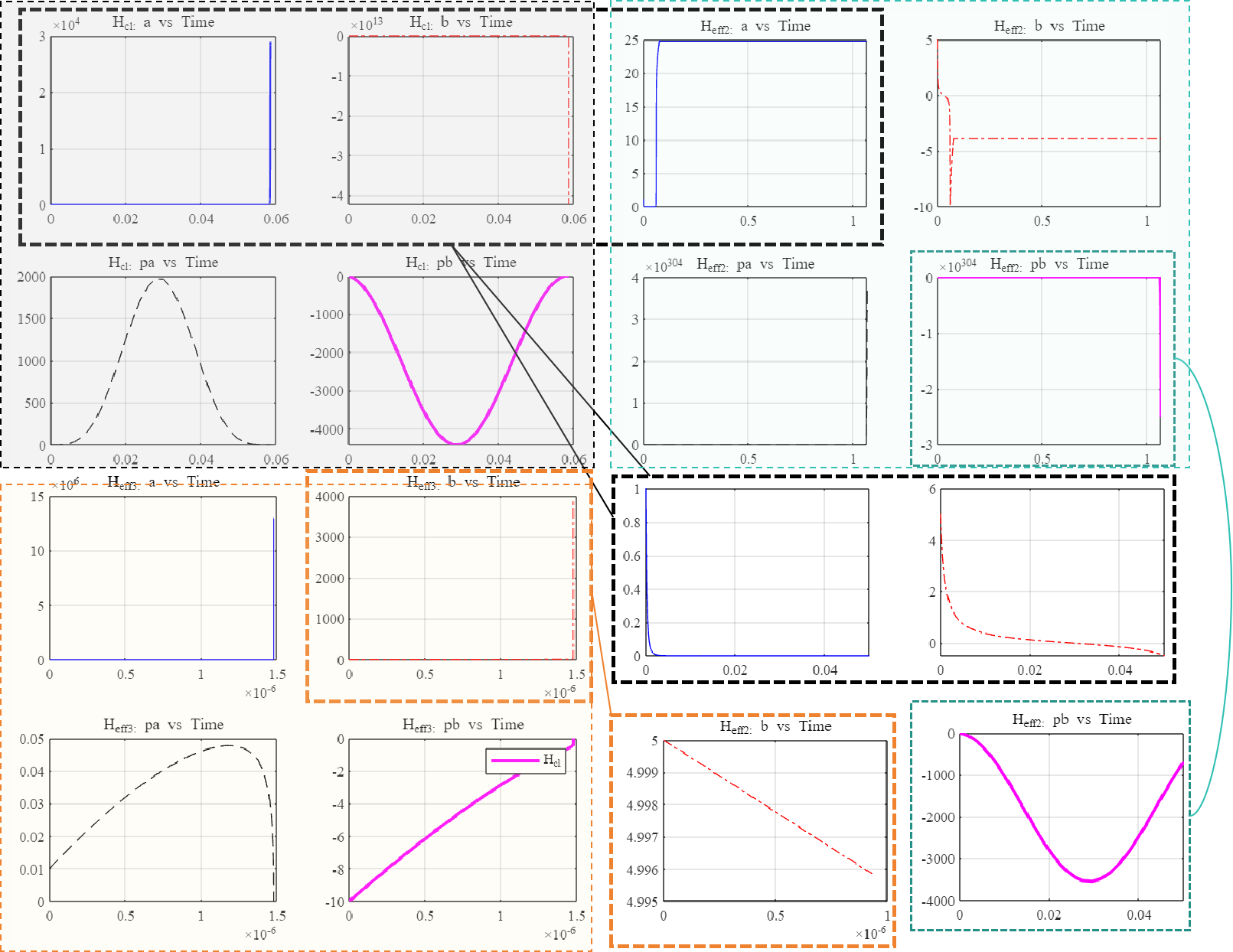}
\caption{It contains three overall parts, which are distinguished by light gray, light cyan and light yellow boxes. Respectively represent the time evolution equation of the Hamiltonian expansion of orders 1-3, and the four sub-figures respectively represent the evolution process of the four parameters. The details at some singular points are enlarged in the lower right corner.}
\label{fig_4}
\end{figure*}

\subsection{Quantum information evolution}

In the following calculations, the initial value conditions for calculation are:

$$x_{p}(0)=0,\quad p_{p}(0)=0$$

Using quantum states to revise the solution of the Schrödinger equation is:

\[\begin{gathered}
  |\psi (u)\rangle  = \cos \theta {e^{i\phi }}\left| {{\phi _0}(u;{x_p}(u),{p_p}(u))} \right\rangle  \hfill \\
   + \sin \theta \left| {{\phi _1}(u;{x_p}(u),{p_p}(u))} \right\rangle  \hfill \\ 
\end{gathered} \]

Just like the model established in the third part, first calculate the evolution of the energy level and the oscillation form according to the calculation results in the previous section, and check whether it tends to the classical solution at the two limits.

\begin{figure*}[!t]
\centering
\includegraphics[width=6in]{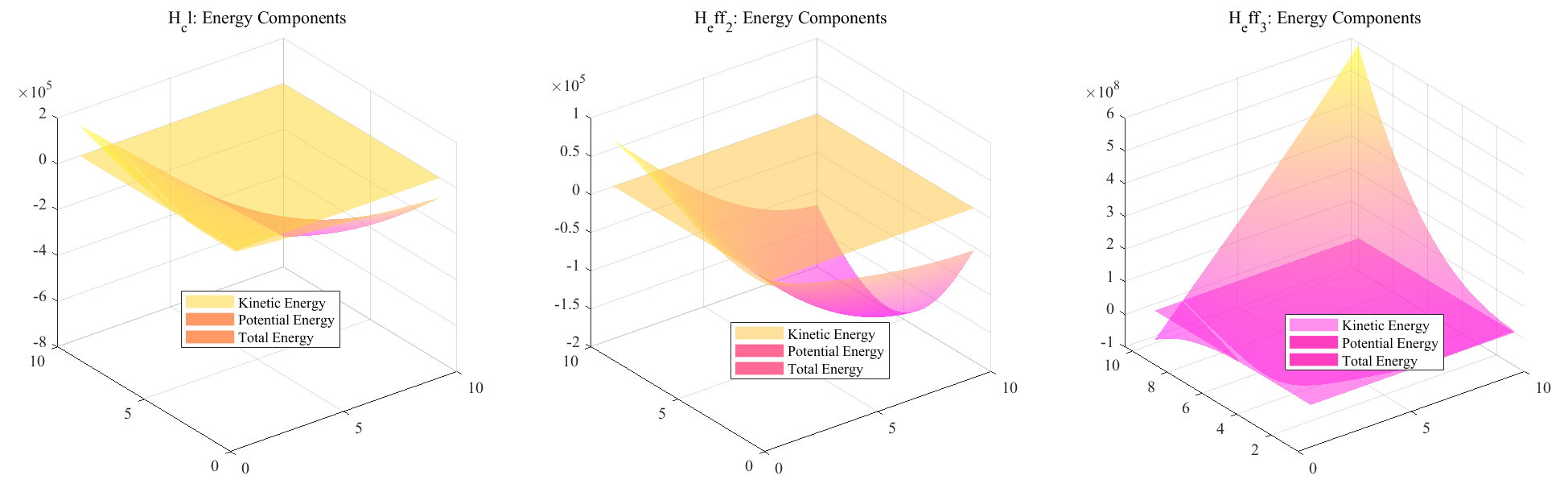}
\caption{The evolution of energy.}
\label{fig_4}
\end{figure*}

Finally, we solve the expectation value evolution equation of the dynamic parameters.

$$
\begin{gathered}
  |{\psi _L}(u;{x_L}(u),{p_L}(u)\rangle  = \cos \theta {e^{i\phi }}|{\phi _0}(u;{x_L}(u),{p_L}(u))\rangle  \hfill \\
   + \sin \theta |{\phi _1}(u;{x_L}(u),{p_L}(u))\rangle  \hfill \\ 
\end{gathered} 
$$

By comparing the evolutions of \(\langle x(u) \rangle\) and \(\langle p(u) \rangle\), it becomes evident that, starting from the third-order expansion, the dynamics confirm the final decoupling process of the quantum system as it propagates within the black hole. This decoupling corresponds to the classical limit dynamics, as under the classical limit, these dynamical parameters ultimately tend toward zero. In the quantum scenario, these energies become coupled with the black hole model, eventually reaching a state of equilibrium. It is essential to note that the late-time state of the quantum oscillator does not constitute an energy eigenstate but rather represents a quantum trajectory. This trajectory can be analyzed from the moment of decoupling, especially when the initial state resides within the quantum bit subspace. Due to the decoupling process at late times, the oscillator should satisfy the free Schrödinger equation, and its energy should ultimately approach a constant value. Concurrently, the time-dependent late-time state should evolve into a constant state. In summary, it is imperative to utilize at least a third-order expansion to establish correlations between the quantum model, energy evolution, and encoding scheme mentioned earlier. This enables the simulation of the information coupling and transmission processes within the black hole system.

\begin{figure*}[!t]
\centering
\includegraphics[width=6in]{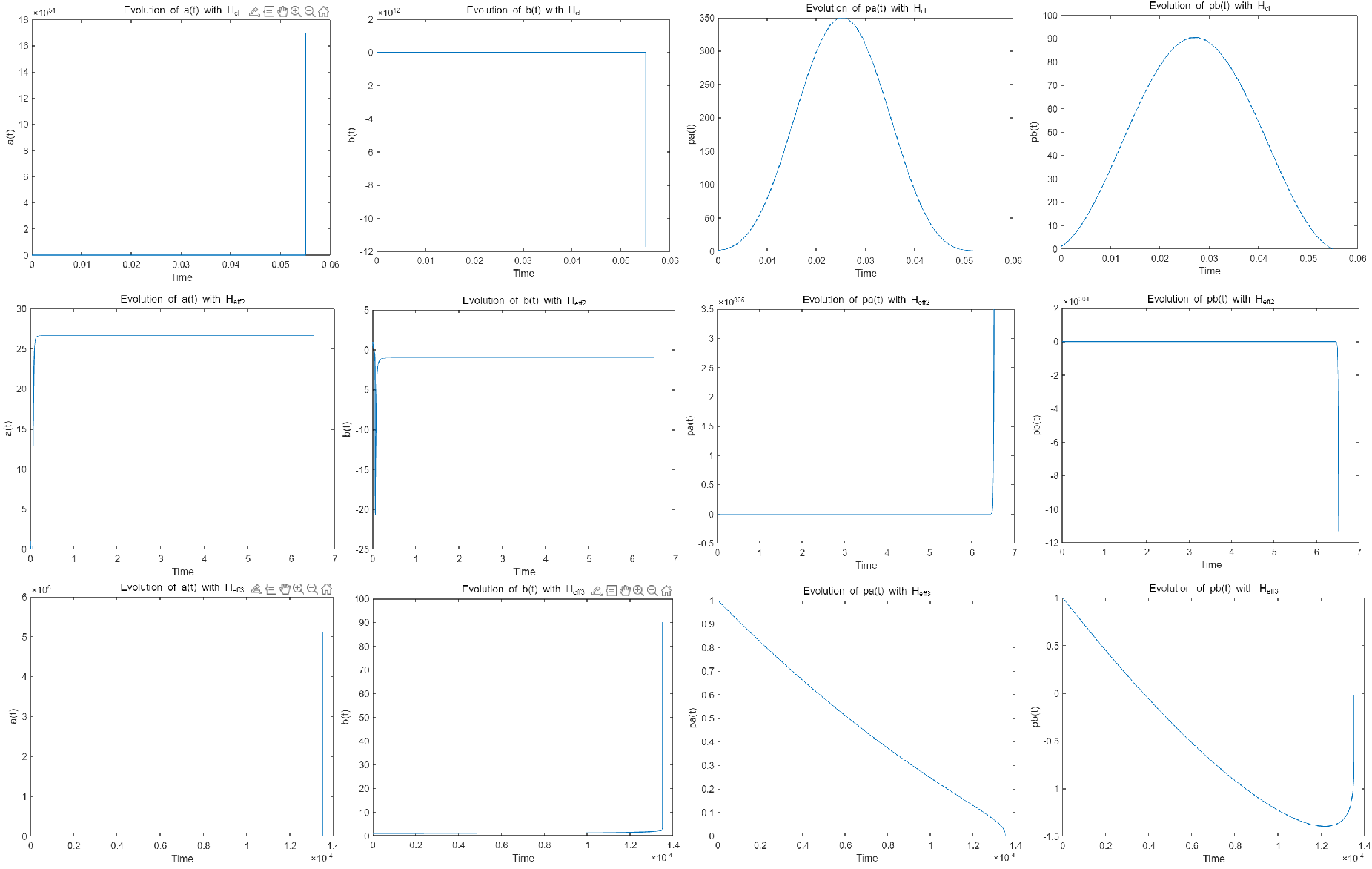}
\caption{kinematic evolution equation}
\label{fig_4}
\end{figure*}

From the comparison results of this numerical calculation, we make a bold guess that the momentum evolution of the analytical expansion of odd-numbered terms is convergent, while the position coordinate evolution of the even-numbered terms expansion is convergent. This further supports the theory about information encoding and information paradox in the model building section. At that time, in order to truly expand this conclusion similar to the incomplete induction method and give a reasonable explanation or solution, we had to carry out a fourth-order or higher expansion, or even provide an analytical solution. This has exceeded the current ability of analytical calculations, and it is necessary to consider developing more accurate numerical calculation methods.

\begin{figure}[!t]
\centering
\includegraphics[width=2.5in]{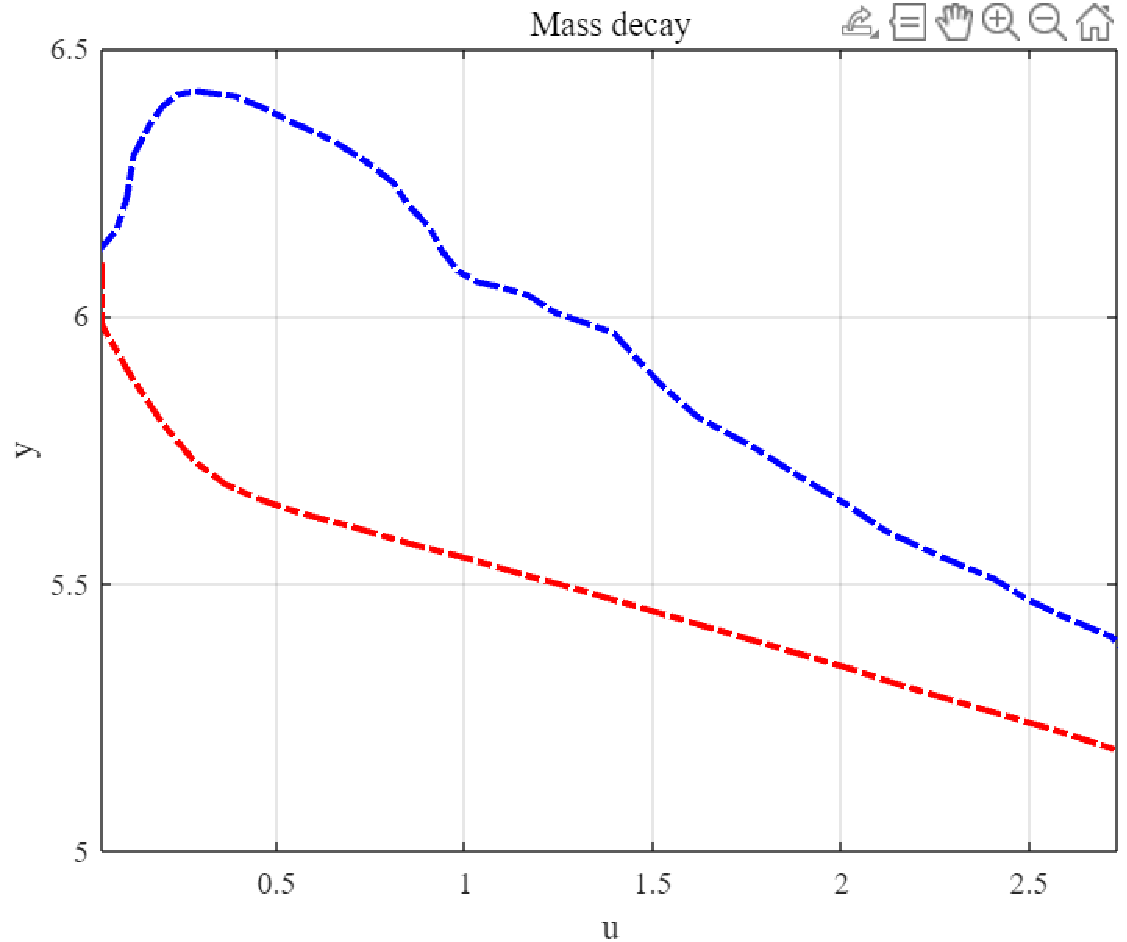}
\caption{mass decay process}
\label{fig_4}
\end{figure}

\section{Conclusion and discussion}

In this paper, we analyze the theoretical model of black holes and the dynamics of information propagation in black holes. We start from the traditional theory of general relativity and establish Hamiltonian models with different analytical accuracy. (It can be seen from the previous description that this is a different order expansion of $\mu$ in the $\mu_0$ method). Compared with traditional theories, our theoretical formula is based on complete loop quantum theory. Our advancement in analytical theory lies in the fact that we do not make too many approximations on the lag function and energy, but strictly analytically expand the Hamiltonian from first order to third order. And use the quantization scheme to simulate the dynamics. Comparing the results of different levels of expansion, it can be concluded that at least three levels of expansion can reasonably simulate the black hole model under the loop quantum universe theory, and can well explain the singularity in the black hole. 

On this basis, we further explore the application of the ring quantum theory model in quantum information dynamics. As mentioned in the introduction, current quantum models (mostly quantum oscillator models) cannot explain quantum information transmission well, mainly because of the existing information paradox (quantum information loss after Hawking radiation). This article uses the theoretical quantum model established previously to further explore quantum information coupling. The mechanisms of encoding and dissemination are discussed, and the computational simulation effects of different analytical precisions (i.e., different order expansions) on the dynamics of information dissemination are discussed.

The final analytical results of this article lead to the conclusion that for the propagation of internal dynamics of black holes and the transmission mechanism of information, at least three levels of analysis can be used to reasonably simulate the propagation of particles and information inside black holes, especially at singularities inside black holes. behavior. This part has been demonstrated by solving the dynamic model of Hamiltonian and the quantum trajectory on Bloch sphere. And quantitatively explained the qualitative explanation of the black hole singularity behavior by loop quantum mechanics.

However, it can be seen from the final research results of this article that if we want to describe the dynamics inside the black hole more completely and accurately, we should still focus on finding analytical solutions or more efficient numerical solution methods. Of course, the analytical expansion method used in this article is also an idea. However, although the third-order analytical expansion has achieved semi-quantitatively satisfactory results, further development is needed to summarize a more complete theory. This was already mentioned in Section IV, which is currently beyond the depth of numerical calculations. In this way, future theoretical work should be focused on developing more efficient and faster calculation methods to provide stronger support for the above theory. Other theoretical incompleteness properties are listed below:

(1)This article does not explore the theoretical analysis results and numerical analysis results established in this article to conduct an in-depth discussion and comparison of the decoding process of quantum information in black holes. Because the current mainstream theory still uses the string theory model. Further consideration needs to be given to the application prospects of the ring quantum theory model.

(2)This article does not consider the many modifications of the Hamiltonian. There are currently many correction schemes for the spherically symmetric Hamiltonian model (some are to simplify calculations, and some take into account more complex quantum corrections). The differences in characteristics of these theoretical models before and after the correction in analyzing black hole propagation dynamics and information dynamics need to be studied.

(3) This article uses the $\mu_0$ method for quantization processing. However, it seems that the $\bar \mu $ method is considered to be more "physical", so whether the calculation results of the modified scheme will obtain better consistent results remains to be studied.

This article reviews the research results of this article from the perspective of improving computational efficiency. Further research plans in the future are still to use theoretically complete analytical formulas, and the complexity of their calculations is considered to rely on quantum computers, using the properties of qubits to provide computational efficiency. Therefore, the research results of this article should be considered as a cross-application and further theoretical attempt in quantum theory, general relativity cosmological model and quantum information transmission.

\appendix
\section{Simplification process of quantum operators}
Using the relation: $\hat{A}\hat{B}=\hat{B}\hat{A}+[\hat{A},\hat{B}]$ Applying the commutator expansion of order $k$ to equation (2.7) we get: 

$$
\begin{aligned}[&\cdots,[[\cdots[[D_{ab}^{\iota}(h_{e}),\hat{p}_{s}^{\alpha_{1}}(e)],\hat{p}_{s}^{\alpha_{2}}(e)]\cdots\quad\times\hat{p}_{s}^{\alpha_{m}}(e)],\hat{p}_{t}^{\alpha_{1}}(e)]\cdots,\hat{p}_{t}^{\alpha_{n}}(e)].\end{aligned}
$$

The result can be derived by:

$$\begin{aligned}[&\cdots[[D_{ab}^{\iota}(h_e),\hat{p}_s^{\alpha_1}(e)],\hat{p}_s^{\alpha_2}(e)]\cdots\hat{p}_s^{\alpha_m}(e)]\\&=(-it)^mD_{aa_1}^{\prime n}(\tau^{\alpha_1})D_{a_1a_2}^{\iota^{\prime}}(\tau^{\alpha_2})\cdots D_{a_{m-1}a_m}^{\prime n}(\tau^{\alpha_m})D_{a_mb}^{\iota}(h_e)\end{aligned}$$

with: $\displaystyle a_{k}=a-\sum_{i=1}^{k}\alpha_{i}$,and:

$$\begin{aligned}
&\cdots[[D_{ab}^{l}(h_{e}),\hat{p}_{t}^{\alpha_{1}}(e)],\hat{p}_{t}^{\alpha_{2}}(e)]\cdots\hat{p}_{t}^{\alpha_{m}}(e)] \\
&=(it)^{m}D_{ab_{m}}^{\prime}(h_{e})D_{b_{m}b_{m-1}}^{\prime\prime}(\tau^{\alpha_{m}})D_{b_{m-1}b_{m-2}}^{\prime\prime}(\tau^{\alpha_{m-1}})\cdots D_{b_{1}b}^{\prime\prime}(\tau^{\alpha_{1}})
\end{aligned}$$

with:$\displaystyle b_k=\sum_{i=1}^kb_i+b$, Continuing the calculation we get:

\[\begin{gathered}
  D_{ab}^\iota ({h_e})\left( {\prod\limits_{i = 1}^m {\hat p_s^{{\alpha _i}}} (e)} \right)\left( {\prod\limits_{j = 1}^n {\hat p_t^{{\beta _j}}} (e)} \right) \hfill \\
   = \left( {\prod\limits_{i = 1}^m {\hat p_s^{{\alpha _i}}} (e)} \right)\left( {\prod\limits_{j = 1}^n {\hat p_t^{{\beta _j}}} (e)} \right)D_{ab}^\iota ({h_e}) - it\sum\limits_{k = 1}^m {\left( {\prod\limits_{i \ne k} {\hat p_s^{{\alpha _i}}} (e)} \right)} \left( {\prod\limits_{j = 1}^n {\hat p_t^{{\beta _j}}} (e)} \right)D_{ac}^{\iota '}({\tau ^{{\alpha _k}}})D_{cb}^\iota ({h_e}) \hfill \\
   + {( - it)^2}\sum\limits_{k < l} ( \prod\limits_{i \notin \{ k,l\} } {\hat p_s^{{\alpha _i}}} (e))(\prod\limits_{j = 1}^n {\hat p_t^{{\beta _j}}} (e))D_{ac}^{\iota '}({\tau ^{{\alpha _k}}})D_{cd}^{\iota '}({\tau ^{{\alpha _l}}})D_{db}^\iota ({h_e}) \hfill \\
   + it\sum\limits_{k = 1}^n {\left( {\prod\limits_{i = 1}^m {\hat p_s^{{\alpha _i}}} (e)} \right)} \left( {\prod\limits_{j \ne k} {\hat p_t^{{\beta _j}}} (e)} \right)D_{ac}^\iota ({h_e})D_{cb}^{\iota '}({\tau ^{{\beta _k}}}) \hfill \\
   + {(it)^2}\sum\limits_{k < l} {\left( {\prod\limits_{i = 1}^m {\hat p_s^{{\alpha _i}}} (e)} \right)} \left( {\prod\limits_{j \notin \{ k,l\} } {\hat p_t^{{\alpha _i}}} (e)} \right)D_{ac}^\iota ({h_e})D_{cd}^{\iota '}({\tau ^{{\beta _l}}})D_{db}^{\iota '}({\tau ^{{\beta _k}}}) \hfill \\
   - {(it)^2}\sum\limits_{k,l} {\left( {\prod\limits_{i \ne k} {\hat p_s^{{\alpha _i}}} (e)} \right)} \left( {\prod\limits_{j \ne l} {\hat p_t^{{\beta _j}}} (e)} \right)D_{ac}^{\iota '}({\tau ^{{\alpha _k}}})D_{cd}^\iota ({h_e})D_{db}^{\iota '}({\tau ^{{\beta _l}}}) + O({t^3}). \hfill \\ 
\end{gathered} \]

If there is more than one holonomy contained in $\hat{O}$, one can
use this procedure to permute them one by one. Finally, $\hat{O}$
is expressed as summation of terms taking the form:

$$\prod_{k=1}^m\hat{p}_s^{\alpha_i}(e)\prod_{k=1}^n\hat{p}_t^{\alpha_i}(e)\prod_{i=1}^lD_{a_ib_i}^{\iota_i}(h_e)$$

Finally, it can be simplified to the form of the product of the last several operators of equation (2.7).

\section{WKB method to solve quantum harmonic oscillator}

Using the WKB approximation rule:

$$\int_{x_1}^{x_2}p(x)dx=\left(n-\frac{1}{2}\right)\pi\hbar;\quad p(x)=\sqrt{2m\left(E-\frac{1}{2}m\omega^2x^2\right)};$$

With the position of:

$$x_2=-x_1=\frac{1}{\omega}\sqrt{\frac{2E}{m}}.$$

So we can get:

\[\begin{gathered}
  \left( {n - \frac{1}{2}} \right)\pi \hbar  = m\omega \int_{ - {x_2}}^{{x_2}} {\sqrt {\frac{{2E}}{{m{\omega ^2}}} - {x^2}} } dx = 2m\omega \int_0^{{x_2}} {\sqrt {x_2^2 - {x^2}} } dx \hfill \\
   = m\omega \left[ {x\sqrt {x_2^2 - {x^2}}  + x_2^2{{\sin }^{ - 1}}(x/{x_2})} \right]_0^{{x_2}} \hfill \\
   = m\omega x_2^2{\sin ^{ - 1}}(1) = \frac{\pi }{2}m\omega x_2^2 = \frac{\pi }{2}m\omega \frac{{2E}}{{m{\omega ^2}}} = \frac{{\pi E}}{\omega } \hfill \\ 
\end{gathered} \]

Therefore, the energy level formula is:

$$E_{n}=\left(n-\frac{1}{2}\right)\hbar\omega $$

Since the WKB numbering starts with \( n = 1 \), whereas for oscillator states we traditionally start with \( n = 0 \),
letting \( n \rightarrow n + 1 \) converts this to the usual formula \( E_n = \left(n + \frac{1}{2}\right) \hbar \omega \). In this case, the WKB approximation
yields the exact results.

\begin{figure}[!t]
\centering
\includegraphics[width=2.5in]{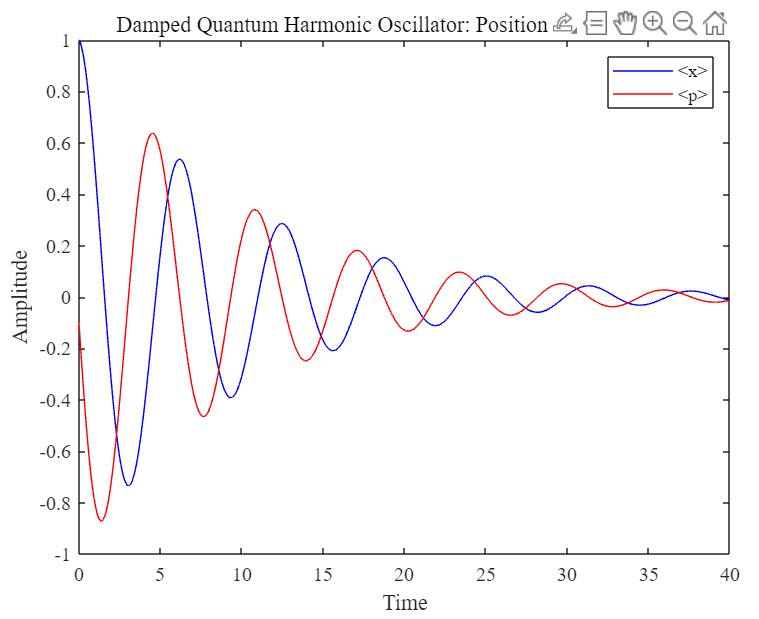}
\caption{The oscillation process of the classical WKB resonator.}
\label{fig_4}
\end{figure}

\end{document}